%% file: paper.tex
\documentstyle[]{mn}
\input{psfig}
\input{macros}
\begin{document}
%

\title[Cosmological Constraints from Clustering \& Lensing]
      {Cosmological Constraints from a Combination of Galaxy Clustering 
         \& Lensing -- I. Theoretical Framework}

\author[van den Bosch et al.]
       {\parbox[t]{\textwidth}{
        Frank C. van den Bosch$^{1}$\thanks{E-mail: frank.vandenbosch@yale.edu}, 
        Surhud More$^{2}$,
        Marcello Cacciato$^{3}$, 
        Houjun Mo$^{4}$, \\
        Xiaohu Yang$^5$} \\ 
           \vspace*{3pt} \\
	$^1$Department of Astronomy, Yale University, PO. Box 208101,
            New Haven, CT 06520-8101\\
	$^2$Kavli Institute for Cosmological Physics, University of
	    Chicago, 933 East 56$^{\rm th}$ Street, Chicago, IL 60637\\
	$^3$Racah Institute of Physics, The Hebrew University,
            Jerusalem 91904, Israel\\
        $^4$Department of Astronomy, University of Massachusetts,
            Amherst MA 01003-9305\\
        $^5$Key Laboratory for Research in Galaxies and Cosmology, Shanghai
            Astronomical Observatory, Nandan Road 80, \\
        $\;$Shanghai 200030, China}


\date{}

\pagerange{\pageref{firstpage}--\pageref{lastpage}}
\pubyear{2012}

\maketitle

\label{firstpage}


\begin{abstract}
  We present a new method that simultaneously solves for cosmology and
  galaxy bias on non-linear scales. The method uses the halo model to
  analytically describe the (non-linear) matter
  distribution, and the conditional luminosity function (CLF) to
  specify the halo occupation statistics. For a given choice of
  cosmological parameters, this model can be used to predict the
  galaxy luminosity function, as well as the two-point correlation
  functions of galaxies, and the galaxy-galaxy lensing signal, both as
  function of scale and luminosity.  These observables have been
  reliably measured from the Sloan Digital Sky Survey.  In this paper,
  the first in a series, we present the detailed, analytical model,
  which we test against mock galaxy redshift surveys constructed from
  high-resolution numerical $N$-body simulations. We demonstrate that
  our model, which includes scale-dependence of the halo bias and a
  proper treatment of halo exclusion, reproduces the 3-dimensional
  galaxy-galaxy correlation and the galaxy-matter cross-correlation
  (which can be projected to predict the observables) with an accuracy
  better than 10 (in most cases 5) percent.
  Ignoring either of these effects, as is often done, results
  in systematic errors that easily exceed 40 percent on scales of
  $\sim 1 h^{-1}\Mpc$, where the data is typically most accurate.
  Finally, since the projected correlation functions of galaxies are
  never obtained by integrating the redshift space correlation
  function along the line-of-sight out to infinity, simply because the
  data only cover a finite volume, they are still affected by residual
  redshift space distortions (RRSDs). Ignoring these, as done in
  numerous studies in the past, results in systematic errors that
  easily exceed 20 perent on large scales ($r_\rmp \gta 10
  h^{-1}\Mpc$).  We show that it is fairly straightforward to correct
  for these RRSDs, to an accuracy better than $\sim 2$ percent, using
  a mildly modified version of the linear Kaiser formalism.
\end{abstract}




\section{Introduction}
\label{sec:intro}

The past decade has seen the emergence of precision cosmology.
Various experiments that probe fluctuations in the cosmic microwave
background (CMB), most notably the Wilkinson Microwave Anisotropy
Probe (WMAP; Bennett \etal 2003) have yielded constraints on various
cosmological parameters at the few percent level (Spergel \etal 2003,
2007; Dunkley \etal 2009; Komatsu \etal 2009, 2011), and ongoing
experiments, such as PLANCK, will tighten these constraints even
further. It is important, though, to complement these data with
non-CMB constraints, such as those provided by supernova Ia, galaxy
clustering, galaxy peculiar velocities, cluster abundances,
gravitational lensing, Lyman $\alpha$ forest and, in the not too
distant future, 21cm tomography of the neutral hydrogen at the era of
reionization. These non-CMB constraints are crucial for (i) breaking
various parameter degeneracies inherent in the CMB data
\footnote{for instance, the CMB as measured by WMAP is consistent with
  a closed Universe with Hubble parameter $h=0.3$ and no cosmological
  constant (e.g. Spergel \etal 2007)}, (ii) constraining certain
cosmological parameters that are largely unconstrained by the CMB,
such as evolution in the equation of state of dark energy, and (iii)
for establishing a true concordance cosmology, i.e., a cosmological
model that is in agreement with all possible data sets.

With the advent of ever larger and more homogeneous galaxy redshift
surveys, such as the Las Campanas Redshift Survey (LCRS; Shectman
\etal 1996), the PSCz (Saunders \etal 2000), the two-Degree Field
Galaxy Redshift Survey (2dFGRS; Colless \etal 2003) and the Sloan
Digital Sky Survey (SDSS; York \etal 2000), there has been a steady
improvement in the tightness and reliability of the corresponding
cosmological constraints. Most of these studies focus on using galaxy
clustering on large scales where one can rely on linear theory. Prime
examples are constraints from (baryon acoustic oscillations in) the
galaxy power spectrum (Percival \etal 2001; Cole \etal 2005;
Eisenstein \etal 2005; Tegmark \etal 2006; H\"utsi 2006; Percival
\etal 2007a,b,c; Padmanabhan \etal 2007; Gaztanaga, Cabr\'e \& Hui
2009; Percival \etal 2010; Blake \etal 2011; Anderson \etal 2012).

However, recently it has also become feasible to accurately model
galaxy clustering on small, non-linear scales using the halo model
approach combined with halo occupation statistics. The halo model
postulates that all dark matter is partitioned over dark matter
haloes, and describes the dark matter density distribution in terms of
the halo building blocks (e.g., Neyman \& Scott 1952; Seljak 2000; Ma
\& Fry 2000; Scoccimarro et al. 2001; Cooray \& Sheth 2002). When
combined with a model that describes how galaxies with certain
properties are distributed over dark matter haloes of different mass,
this can be used to make predictions for the clustering properties of
galaxies on all scales that are observationally accessible (e.g.,
Jing, Mo \& B\"orner 1998; Berlind \& Weinberg 2002; Cooray \& Sheth
2002; Yang, Mo \& van den Bosch 2003).

This approach has been used extensively in recent years to constrain
the galaxy-dark matter connection, i.e., the connection between galaxy
properties and halo mass, which holds important information regarding
galaxy formation. On large, linear scales, the two-point correlation
function between haloes of mass $M$ can be written as $\xi_{\rm
  hh}(r|M) = b^2_\rmh(M) \, \xi^{\rm lin}_{\rm mm}(r)$, with $\xi^{\rm
  lin}_{\rm mm}(r)$ the two-point correlation function of the linear
matter distribution and $b_\rmh(M)$ the linear halo bias (e.g., Mo \&
White 1996). Similarly, for galaxies of a given luminosity, one has
that $\xi_{\rm gg}(r|L) = b^2_\rmg(L) \, \xi^{\rm lin}_{\rm mm}(r)$,
with $b_\rmg(L)$ the bias of galaxies of luminosity $L$. Hence, one
can use $\xi_{\rm gg}(r|L)$ to infer the {\it average} mass of haloes
that host galaxies of luminosity $L$ by simply finding the $M$ for
which $b_\rmh(M) = \left[\xi_{\rm gg}(r|L)/\xi^{\rm lin}_{\rm
    mm}(r)\right]^{1/2}$. By comparing the observed abundance of
galaxies of luminosity $L$ to the predicted abundance of haloes of
mass $M$, one subsequently infers the average number of galaxies per
halo.  Hence, measurements of $\xi_{\rm gg}(r|L)$ can be used to
constrain halo occupation statistics, and this technique has been
widely used (Jing et al. 1998, 2002; Peacock \& Smith 2000; Bullock,
Wechsler \& Somerville 2002; Magliocchetti \& Porciani 2003; Yang
\etal 2003, 2004; van den Bosch \etal 2003a, 2007; Porciani,
Magliocchetti \& Norberg 2004; Wang \etal 2004; Zehavi \etal 2004,
2005; Zheng 2004; Abazajian \etal 2005; Collister \& Lahav 2005;
Tinker \etal 2005, 2006; Lee \etal 2006). Note, though, that this
method requires knowledge of both $b_\rmh(M)$ and $\xi^{\rm lin}_{\rm
  mm}(r)$, both of which are strongly cosmology
dependent. Consequently, the resulting halo occupation statistics are
also cosmology dependent (see e.g., Zheng \etal 2002; Berlind \&
Weinberg 2002; van den Bosch \etal 2007; Cacciato \etal
2009). Although this makes it difficult to calibrate galaxy formation
models using halo occupation statistics (e.g., Berlind \etal 2003), it
also implies that one can use this method to constrain cosmological
parameters as long as one has some independent constraints on halo
occupation statistics.

Various approaches to constrain cosmological parameters along these
lines have been used in recent years. Abazajian \etal (2005) have
shown that the degeneracy between occupation statistics and cosmology
can (at least partially) be broken by using the correlation function
itself, as long as one includes data on sufficiently small scales
(i.e., the one-halo term).  Using the projected correlation functions
measured from the SDSS and allowing the cosmological parameters to
vary within constraints imposed by various CMB experiments, they were
able to obtain constraints that were significantly tighter than those
from the CMB alone, with $\Omega_\rmm = 0.26 \pm 0.03$ and $\sigma_8 =
0.83 \pm 0.04$.

Zheng \etal (2002) suggested that one can break the degeneracy
between halo occupation model and cosmology by using the peculiar
velocities of galaxies as inferred from the redshift space distortions
in the two-point correlation function. This idea was used by Yang
\etal (2004), who concluded that the power-spectrum normalization,
$\sigma_8$, needs to be of the order of $\sim 0.75$ (assuming
$\Omega_\rmm = 0.3$), significantly lower than the value then
advocated by WMAP. Very similar results were obtained by van den Bosch
\etal (2007) and by Tinker \etal (2007).  The latter used a much more
sophisticated treatment of redshift space distortions developed by
Tinker, Weinberg \& Zheng (2006) and Tinker (2007).

An alternative approach for breaking the degeneracy between halo
occupation model and cosmology is to use constraints on the (average)
mass-to-light ratios of dark matter haloes. This method was first used
by van den Bosch \etal (2003b) and Tinker \etal (2005), who were able
to obtain relatively tight constraints on $\Omega_\rmm$ and $\sigma_8$
from combinations of clustering data plus constraints on the
mass-to-light ratios of clusters. Interestingly, both studies again
found evidence for a relatively low value of the power spectrum
normalization: $\sigma_8 \simeq 0.75$ for $\Omega_\rmm = 0.25$.

Along similar lines, one can also use a combination of clustering and
galaxy-galaxy lensing. The latter effectively probes the galaxy-dark
matter cross correlation, and therefore holds information regarding
the mass-to-light ratios of dark matter haloes covering a wide range
in halo mass. Since its first detection by Brainerd, Blandford \&
Smail (1996), the accuracy of galaxy-galaxy lensing measurements has
increased to the extent of yielding high signal-to-noise ratio
measurements over a significant dynamic range in galaxy luminosity
and/or stellar mass (e.g., Fisher \etal 2000; Hoekstra \etal 2002;
Sheldon \etal 2004, 2009; Mandelbaum \etal 2006, 2009; Leauthaud \etal
2007). Similar to the galaxy-galaxy autocorrelation function, the
galaxy-matter cross correlation function can be accurately
modeled using the halo model (Guzik \& Seljak 2001, 2002; Mandelbaum
\etal 2005; Yoo \etal 2006; Cacciato \etal 2009; Leauthaud \etal 2011,
2012; van Uitert \etal 2011). Hence, the combination of galaxy
clustering and galaxy-galaxy lensing is ideally suited to constrain
cosmological parameters, as demonstrated in detail by Yoo \etal
(2006). A first application of this idea by Seljak \etal (2005), using
the model of Guzik \& Seljak (2002) and the galaxy-galaxy lensing data
of Mandelbaum \etal (2006), combined with WMAP constraints, yielded
$\sigma_8 = 0.88 \pm 0.06$, only marginally consistent with the values
obtained from the cluster mass-to-light ratios and/or the redshift
space distortions mentioned above.  However, more recently, two
different analyses based on the same galaxy-galaxy lensing data by
Cacciato \etal (2009) and Li \etal (2009) both argued that a flat
$\Lambda$CDM cosmology with $(\Omega_\rmm,\sigma_8) = (0.238, 0.734)$
is in much better agreement with the data than a $(0.3,0.9)$ model.
Although the reason for the disagreement between these studies and
that of Seljak \etal (2005) is probably related to the different
modelling approaches, these studies all have demonstrated that a
combination of clustering and lensing data holds great potential for
constraining cosmological parameters.
 
This is the first paper in a series in which we use a combination of
galaxy clustering and galaxy-galaxy lensing data to constrain
cosmological parameters. In this paper we present the theoretical
framework and test the accuracy of our method using mock data.  In
More \etal 2012a (hereafter Paper~II) we present a Fisher matrix
analysis to identify parameter-degeneracies and to assess the accuracy
with which various cosmological parameters can be constrained using
the methodology presented here. Finally, in Cacciato \etal 2012b
(hereafter Paper~III) we apply our analysis to the actual SDSS data to
constrain cosmological parameters (in particular $\Omega_\rmm$ and
$\sigma_8$) under the assumption of a `standard' flat $\Lambda$CDM
cosmology.

Throughout this paper, unless specifically stated otherwise, all radii
and densities will be in comoving units, and log is used to refer to
the 10-based logarithm. Quantities that depend on the Hubble parameter
will be written in units of $h = H_0/(100 \kmsmpc)$.
 

\section{Model Description} 
\label{sec:model}

Our main goal is to use galaxy clustering and galaxy-galaxy lensing,
measured as function of luminosity from the main galaxy sample in the
SDSS, to simultaneously constrain cosmology and halo occupation
statistics. As detailed in papers II and III, the data that we will use
consists of (i) the galaxy luminosity function, $\Phi(L,z)$, at the
median redshift of the SDSS main galaxy sample ($z \simeq 0.1$), (ii)
the projected two-point correlation functions, $w_\rmp(r_\rmp|L_1,L_2,z)$,
for galaxies in six luminosity bins, $[L_1,L_2]$, each with its own
median redshift $z$, and (iii) the corresponding excess surface
densities (ESD), $\Delta\Sigma(R|L_1,L_2,z)$.  

The projected correlation function, $w_\rmp(r_\rmp|L_1,L_2,z)$, is
related to the corresponding galaxy-galaxy correlation function in
real space, $\xi(r|L_1,L_2,z)$, via a simple Abel integral
\begin{equation}
  w_\rmp(r_\rmp|L_1,L_2,z) = 2 \int_{r_\rmp}^{\infty} \xi_{\rm gg}(r|L_1,L_2,z) 
  {r \, {\rm d}r \over \sqrt{r^2 - r^2_\rmp}}\,,
\end{equation}
(but see \S\ref{sec:proj} below).  The ESD,
$\Delta\Sigma(R|L_1,L_2,z)$, is related to the tangential shear,
$\gamma_\rmt(R|L_1,L_2,z)$, measured around galaxies (the lenses) at
redshift $z$ with luminosities in the range $[L_1,L_2]$ according to
\begin{eqnarray}\label{shear} 
\Delta\Sigma (R|L_1,L_2,z) & = & \overline{\Sigma}(<R|L_1,L_2,z) - 
\Sigma(R|L_1,L_2,z) \nonumber \\
& = & \gamma_{\rm t}(R|L_1,L_2,z) \Sigma_{\rm crit} \,.
\end{eqnarray}
Here $\Sigma_{\rm crit}$ is a geometrical parameter that depends on
the comoving distances of the sources and lenses,
$\Sigma(R|L_1,L_2,z)$ is the azimuthally-averaged projected surface
mass density of the gravitational lenses, which is related to the
galaxy-matter cross correlation function, $\xi_{\rm
  gm}(r|L_1,L_2,z)$, according to
\begin{eqnarray}\label{Sigma_approx}
\lefteqn{\Sigma(R|L_1,L_2,z) = 2\,\bar{\rho}_\rmm(z)} \nonumber \\
 & & \int_{R}^{\infty} \left[1+\xi_{\rm gm}(r|L_1,L_2,z)\right] \, 
{ r \, \rmd r \over \sqrt{r^2 - R^2}}\,,
\end{eqnarray}
and $\overline{\Sigma}(<R|L_1,L_2,z)$ is its average inside $R$;
\begin{equation}\label{averageSigma}
\bar{\Sigma}(<R|L_1,L_2,z)  = {2\over R^2}
\int_0^R \Sigma(R'|L_1,L_2,z) \, R' \, \rmd R'\,,
\end{equation}
(Miralda-Escud\'e 1991; Sheldon \etal 2004; see also \S\ref{sec:xigg}).  

In this section, we present analytical expressions for
$w_\rmp(r_\rmp|L_1,L_2,z)$, $\Delta\Sigma(R|L_1,L_2,z)$ and
$\Phi(L,z)$.  For completeness and clarity we present a detailed,
step-by-step derivation of our method, and we will emphasize where it
differs from that of previous authors.  The backbone of our model is
the halo model, in which the matter distribution in the Universe is
described in terms of its halo building blocks (see Cooray \& Sheth
2002 and Mo, van den Bosch \& White 2010 for comprehensive reviews).
After a detailed description of how the halo model can be used to
compute the power spectrum of the dark matter mass distribution
(\S\ref{sec:halomodel}), we show how the halo model can be
complemented with a model for halo occupation statistics which allows
one to compute $w_\rmp(r_\rmp|L_1,L_2,z)$, $\Delta\Sigma(R|L_1,L_2,z)$
and $\Phi(L,z)$ for a given cosmology.

In order to keep the derivations concise, in what follows we will not
explicitly write down the dependencies on $L_1$ and $L_2$.

\subsection{The halo model}
\label{sec:halomodel}

Throughout this paper we define dark matter haloes as spherical
overdensity regions with a radius, $r_{200}$, inside of which the
average density is 200 times the average density of the
Universe. Hence, the mass of a halo is
\begin{equation}
M = {4 \pi \over 3} \, 200 \, \bar{\rho}_\rmm \, r^{3}_{200} \,.
\end{equation}

Under the assumption that all dark matter is bound in virialized dark
matter haloes, the density perturbation field at redshift $z$, defined
by
\begin{equation}
\delta_\rmm(\bx,z) \equiv 
{\rho_\rmm(\bx,z) \over \bar{\rho}_\rmm} - 1\,,
\end{equation}
can be written in terms of the spatial distribution of dark matter
haloes and their internal density profiles.  Throughout we assume that
dark matter haloes are spherically symmetric and have a density
profile, $\rho_\rmh(r|M,z) = M \, u_\rmh(r|M,z)$, that depends only on
mass, $M$, and redshift, $z$.  Note that $\int u_\rmh(\bx|M,z) \,
\rmd^3 \bx = 1$.

Now imagine  that space  (at some redshift  $z$) is divided  into many
small volumes,  $\Delta V_i$, which are  so small that  none of them
contains more than one halo center.  The occupation number of haloes
in the $i^{\rm th}$ volume,  ${\cal N}_{\rmh, i}$, is therefore either
0   or  1,   and   so   $\calN_{\rmh,  i}   =   \calN^2_{\rmh,  i}   =
\calN^3_{\rmh,i}...$.   In  terms  of  these  occupation  numbers  the
density field of the (dark) matter can formally be written as
\begin{equation}\label{rhodm}
\rho_\rmm(\bx,z) = \sum_{i} \calN_{\rmh, i} \, M_i \,
u_\rmh(\bx - \bx_{i}|M_{i},z) \, ,
\end{equation} 
where $M_i$ is  the mass of the halo whose center  is in $\Delta V_i$.
Using that the ensemble  average $\langle \calN_{\rmh, i} \,
M_i \, u_\rmh(\bx - \bx_{i}|M_{i},z)\rangle$  is equal to $\int \rmd M
\,  n(M,z) \,  M  \, \Delta  V_i  \, u_\rmh(\bx  - \bx_i|M,z)$,  where
$n(M,z)$ is the halo mass function, we have that
\begin{eqnarray}
\langle \rho_\rmm(\bx,z) \rangle & = & 
\int \rmd M \, M \, n(M,z) \sum_i \Delta V_i \, u_\rmh(\bx - \bx_i|M,z) \nonumber \\
& = &  \int \rmd M \, M \, n(M,z) \int \rmd^3\bx' \, u_\rmh(\bx - \bx'|M,z) \nonumber \\
& = & \bar{\rho}_\rmm \,,
\end{eqnarray}
where the last equality follows from the normalization of
$u_\rmh(\bx|M,z)$ and from the halo model ansatz that all dark matter
is partitioned over dark matter haloes.

Similar to $\delta_\rmm$ we can also define the halo density contrast
$\delta_\rmh$.  Ignoring possible stochasticity in the relation
between $\delta_\rmm$ and $\delta_\rmh$, we can use a Taylor series
expansion to write
\begin{equation}\label{hmfield}
\delta_\rmh(\bx;M,z) = \delta_\rmh(\delta_\rmm) =
\sum_{n=0}^{\infty} {b_{\rmh, n}(M,z) \over n!} \, \delta^n_\rmm(\bx,z)\,,
\end{equation}
(Fry \& Gaztanaga 1993; Mo, Jing \& White 1997), where $b_{\rmh, n}$
is called the halo bias factor of order $n$. Although the requirement
that $\langle \delta_\rmh \rangle=0$ implies that $b_{\rmh, 0} =
-\sum_{n=2}^{\infty} b_{\rmh, n} \langle \delta^n_\rmm \rangle/n!$,
which in general is not zero, one can ignore $b_{\rmh, 0}$ since in
Fourier space it only contributes to the galaxy power spectrum for
wavevector $\bk =0$.  Furthermore, on large scales we have that
$|\delta_\rmm| \ll 1$, so that we can also neglect the higher-order
($n>1$) bias factors. Hence, on large scales the cross correlation
function of haloes of mass $M_1$ and haloes of mass $M_2$ can be
written as
\begin{equation}\label{xihhdef}
\xi_{\rm hh}(r|M_1,M_2,z) \simeq b_\rmh(M_1,z) \, b_\rmh(M_2,z) \,
\xi^{\rm lin}_{\rm mm}(r,z)\,,
\end{equation}
where $\xi^{\rm lin}_{\rm mm}(r,z)$ is the two-point correlation
function of the initial density perturbation field, linearly
extrapolated to redshift $z$, and we have used $b_\rmh(M,z)$ as
shorthand notation for the linear halo bias $b_{\rmh, 1}(M,z)$. One
can extend this prescription to the mildly non-linear regime, in which
one can no longer ignore the higher-order bias terms, by replacing
$\xi^{\rm lin}_{\rm mm}(r,z)$ with the {\it non-linear} two-point
matter correlation function, $\xi_{\rm mm}(r,z)$, and by including a
radial dependence of the halo bias, $\zeta(r,z)$ (which effectively
captures the effect of the higher-order bias parameters, see
\S\ref{sec:radbiasfunction} below). Under the assumption that haloes
are spherical, one then obtains that
\begin{eqnarray}\label{xihh}
\lefteqn{1 + \xi_{\rm hh}(r|M_1,M_2,z) = } \\
& & \left[ 1 + b_\rmh(M_1,z) b_\rmh(M_2,z) \zeta(r,z) 
\xi_{\rm mm}(r,z) \right] \, \Theta(r-r_{\rm min})\,, \nonumber 
\end{eqnarray}
where $\Theta(x)$ is the Heaviside step function, which assures that
$\xi_{\rm hh}(r,z|M_1,M_2) = -1$ for $r < r_{\rm min}$ in order to
account for halo exclusion, i.e., the fact that dark matter haloes
cannot overlap. In principle, one expects that $r_{\rm min} = r_{\rm
  min}(M_1,M_2,z) = r_{200}(M_1,z) + r_{200}(M_2,z)$. However, the
halo finder used by Tinker \etal (2008), whose halo mass function we
use, {\it does} allow overlap of haloes in that any halo is considered
a host halo as long as its center does not lie within the outer radius
of another halo. Therefore, to be consistent, we follow Tinker \etal
(2012) and Leauthaud \etal (2011), and adopt that $r_{\rm min} = {\rm
  MAX}\left[r_{200}(M_1,z),r_{200}(M_2,z)\right]$.

For computational  convenience, we will  be working in  Fourier space.
To that extent we define the Fourier transform of $\rho_\rmm(\bx,z)$ as
\begin{eqnarray}\label{rhoFT}
\tilde{\rho}_\rmm(\bk,z) & \equiv & {1 \over V} \int \rho_\rmm(\bx,z) 
       \rme^{-i \bk\cdot\bx} \rmd^3 \bx \nonumber \\
 & = & {1 \over V} \sum_i \calN_{\rmh, i} M_i \,
        \tilde{u}_\rmh(\bk|M_i,z) \, \rme^{-i\bk\cdot\bx_i}\,,
\end{eqnarray}
where $V$ is the volume over which the Universe is assumed to be
periodic, and
\begin{equation}
\tilde{u}_\rmh(\bk|M,z) \equiv \int u(\bx|M,z) \,
\rme^{-i\bk\cdot\bx} \rmd^3\bx\,,
\end{equation}
is the Fourier transform of the normalized halo density profile.  With
our definition of the Fourier transform, the (non-linear)
matter-matter power spectrum is defined as
\begin{eqnarray}\label{Pkdef}
P_{\rmm\rmm}(\bk,z) & = & V \langle |\delta_\rmm(k)|^2 \rangle \nonumber \\
& = & {V \over \bar{\rho}^2_\rmm}
\langle \tilde{\rho}_\rmm(\bk,z) \tilde{\rho}_\rmm^{\ast}(\bk,z) \rangle -  
V \delta^\rmD(\bk)\,,
\end{eqnarray}
where 
\begin{equation}
\delta^\rmD(\bk) = {1 \over V} \int \rme^{-i \bk\cdot\bx} \rmd^3\bx\,,
\end{equation}
is the Dirac delta function, $\rho^{\ast}$ indicates the complex
conjugate of $\rho$, and we have used that $\tilde{\rho}_\rmm(0) =
\bar{\rho}_\rmm$.

Using Eq.~(\ref{rhoFT}) we have that
\begin{eqnarray}
\lefteqn{\langle \tilde{\rho}_\rmm(\bk,z) \tilde{\rho}_\rmm^{\ast}(\bk,z) \rangle 
= {1 \over V^2} \sum_i \sum_j } \\
& & \langle \calN_{\rmh, i} M_i
\calN_{\rmh, j} M_j \tilde{u}_\rmh(\bk|M_i,z)
\tilde{u}^{\ast}_\rmh(\bk|M_j,z) \rme^{-i \bk\cdot(\bx_i - \bx_j)}
\rangle \nonumber \,,
\end{eqnarray}
which we split in two terms: the one-halo term, for which $j=i$, and
the two-halo term with $j\ne i$. The former can be written as
\begin{eqnarray}
\lefteqn{\langle \tilde{\rho}_\rmm(\bk,z) \tilde{\rho}_\rmm^{\ast}(\bk,z)
\rangle^{\rm 1h} = {1 \over V^2} \sum_i \langle \calN_{\rmh, i} M^2_i
|\tilde{u}_\rmh(\bk|M_i,z)|^2 \rangle} \nonumber \\
& = & {1 \over V} \int \rmd M \, M^2 \, n(M,z) \,
|\tilde{u}_\rmh(\bk|M,z)|^2 \,,
\end{eqnarray}
where we have used that $\calN^2_{\rmh, i} = \calN_{\rmh, i}$. For the
2-halo term  we use the fact that  we are free to  choose $\Delta V_i$
arbitrary small, so that
\begin{eqnarray}\label{Pkdm2h}
\lefteqn{\langle \tilde{\rho}_\rmm(\bk,z) \tilde{\rho}_\rmm^{\ast}(\bk,z) 
\rangle^{\rm 2h} = {1 \over V^2} \int \rmd^3\by_1 \int\rmd^3\by_2} \nonumber \\
& & \int \rmd M_1 \, M_1 \, n(M_1,z) \, \tilde{u}_\rmh(\bk|M_1,z) \nonumber \\
& & \int \rmd M_2 \, M_2 \, n(M_2,z) \, \tilde{u}^{\ast}_\rmh(\bk|M_2,z) \nonumber \\
& & \left[ 1 + \xi_{\rm hh}(\by_1 - \by_2,z|M_1,M_2) \right] \rme^{-i \bk\cdot(\by_1 - \by_2)}\,.
\end{eqnarray}
Here we have accounted for the fact that dark matter haloes are
clustered, as described by the two-point halo-halo correlation
function $\xi_{\rm hh}(\br,z|M_1,M_2)$. 

Hence, using Eq.~(\ref{xihh}), which properly accounts for halo
exclusion, we have that
\begin{eqnarray}\label{Pkdm2hb}
\lefteqn{\langle \tilde{\rho}_\rmm(\bk,z) \tilde{\rho}_\rmm^{\ast}(\bk,z) 
\rangle^{\rm 2h} = {1 \over V} 
\int \rmd M_1 \, M_1 \, n(M_1,z) \, \tilde{u}_\rmh(\bk|M_1,z) } \nonumber \\
& & \int \rmd M_2 \, M_2 \, n(M_2,z) \, \tilde{u}_\rmh(\bk|M_2,z)
Q(k|M_1,M_2,z)\,.
\end{eqnarray}
Here we have used that $\tilde{u}^{\ast}(\bk|M,z) =
\tilde{u}(\bk|M,z)$, which follows from the fact that $u(\bx|M,z)$ is
real and even, and we have defined
\begin{eqnarray}\label{QkM}
\lefteqn{Q(k|M_1,M_2,z) \equiv } \nonumber \\
& & 4 \pi \int_{r_{\rm min}}^{\infty}  \left[ 1+
\xi_{\rm hh}(r,z|M_1,M_2)\right] \,{\sin kr \over kr}\, r^2 \,\rmd r\,,
\end{eqnarray}
with $k = |\bk|$ and with $\xi_{\rm hh}(r,z|M_1,M_2)$ given by
Eq.~(\ref{xihh}).

Combining Eqs.~(\ref{Pkdef})-(\ref{Pkdm2hb}), and using that haloes
are defined to be spherically symmetric, we finally obtain that
\begin{equation}
P_{\rm mm}(\bk,z) = P^{\rm 1h}_{\rmm\rmm}(k,z) + P^{\rm 2h}_{\rmm\rmm}(k,z)
- V \delta^\rmD(\bk)\,,
\end{equation}
where 
\begin{equation}
P^{\rm 1h}_{\rmm\rmm}(k,z) = {1 \over \bar{\rho}^2_\rmm} \int\rmd M
\, M^2 \, n(M,z) \,|\tilde{u}_\rmh(k|M,z)|^2\,,
\end{equation}
and
\begin{eqnarray}\label{P2hexc}
\lefteqn{P^{\rm 2h}_{\rmm\rmm}(k,z) = {1 \over \bar{\rho}^2_\rmm} 
\int \rmd M_1 \, M_1 \, n(M_1,z) \, \tilde{u}_\rmh(k|M_1,z)} \nonumber \\
& & \int \rmd M_2 \, M_2 \, n(M_2,z) \, \tilde{u}_\rmh(k|M_2,z)
Q(k|M_1,M_2,z)\,.
\end{eqnarray}
Our treatment of halo exclusion is similar to that of Smith,
Scoccimarro \& Sheth (2007) and Smith, Desjacques \& Marian (2011),
except that we have included the (semi-empirical) factor $\zeta(r,z)$
to account for the radial dependence of halo bias.  As shown in Smith
\etal (2011), Eq.~(\ref{P2hexc}) has the correct asymptotic behavior
at both large and small scales. This is an important improvement over
a number of {\it approximate} methods that have been advocated and
which typically involve adopting an upper limit for the mass interval
used in the integral for the 2-halo term of the power spectrum (e.g.,
Takada \& Jain 2003; Zheng 2004; Abazajian \etal 2005; Tinker \etal
2005, 2012; Yoo \etal 2006; Leauthaud \etal 2011). None of these
methods, however, are mathematically correct. Furthermore, accurate,
numerical evaluation of Eq.~(\ref{P2hexc}) is not significantly more
CPU demanding than using the approximate method, largely rescinding
its main motivation. Finally, as shown in Smith \etal (2011),
Eq.~(\ref{P2hexc}) has the additional advantage that it appears to
resolve the well-known problem of excess large-scale power in the halo
model. This problem arises due to the fact that the 1-halo term
approaches a constant value on large scales in Fourier space,
significantly in excess of the shot noise (see discussions in Cooray
\& Sheth 2002; Smith \etal 2003; Crocce \& Scoccimarro 2008). A proper
treatment of halo exclusion, as adopted here, (almost) nullifies this
large scale power of the 1-halo term.

\subsection{The galaxy-galaxy correlation function}
\label{sec:xigg}

If one assumes that each galaxy resides in a dark matter halo, the
halo model described above can also be used to compute the
galaxy-galaxy correlation function or the galaxy-matter cross
correlation function. All that is needed is a statistical description
of how galaxies are distributed over dark matter haloes of different
mass.  To that extent we use the conditional luminosity function
(hereafter CLF) introduced by Yang \etal (2003).  The CLF, $\Phi(L|M)
\rmd L$, specifies the {\it average} number of galaxies with
luminosities in the range $L \pm \rmd L/2$ that reside in a halo of
mass $M$.

Throughout we ignore a potential redshift dependence of the CLF. Since
the data that we use to constrain the CLF only covers a narrow range
in redshift (see Paper~III), this assumption will not have a strong
impact on our results.  Once the CLF is specified, the galaxy
luminosity function at redshift $z$, $\Phi(L,z)$, simply follows from
integrating over the halo mass function, $n(M,z)$;
\begin{equation}
\Phi(L,z) = \int \Phi(L|M) \, n(M,z) \, \rmd M\,.
\end{equation}
In what follows, we will always be concerned with galaxies in a
specific luminosity interval $[L_1,L_2]$. The average number density
of such galaxies follows from the CLF according to
\begin{equation}\label{avgngal}
\bar{n}_\rmg(z) = \int \langle N_\rmg|M \rangle \, n(M,z) \, \rmd M\,,
\end{equation}
where
\begin{equation}\label{avgnm}
\langle N_\rmg|M\rangle = \int_{L_1}^{L_2} \Phi(L|M) \rmd L\,,
\end{equation}
is the average number of galaxies with $L_1 < L < L_2$ that reside in
a halo of mass $M$.

For  reasons  that  will  become  clear below,  we  split  the  galaxy
population in centrals  (defined as those galaxies that  reside at the
center of their  host halo) and satellites (those  that orbit around a
central), and we split the CLF in two terms accordingly:
\begin{equation}
\Phi(L|M) = \Phi_\rmc(L|M) + \Phi_\rms(L|M)\, ,
\end{equation}
where $\Phi_\rmc(L|M)$ and $\Phi_\rms(L|M)$ represent central and
satellite galaxies, respectively (cf., Cooray \& Milosavljevic 2005).
Similarly, we write the number density of galaxies, $n_\rmg(\bx,z)$,
as the sum of the contribution of centrals, $n_\rmc(\bx,z)$, and that
of satellites, $n_\rms(\bx,z)$, so that
\begin{eqnarray}\label{ngalsplit}
\delta_\rmg(\bx,z) & \equiv & {n_\rmg(\bx,z) - \bar{n}_\rmg(z) \over
\bar{n}_\rmg(z)} \nonumber \\
& = & f_\rmc(z) \delta_\rmc(\bx,z) + f_\rms(z) \delta_\rms(\bx,z)\,.
\end{eqnarray}
Here $f_\rmc(z)=\bar{n}_\rmc(z)/\bar{n}_\rmg(z)$ is the central
fraction, $f_\rms(z)=\bar{n}_\rms(z)/\bar{n}_\rmg(z)=1-f_\rmc(z)$ is
the satellite fraction, and $\delta_\rmc(\bx,z)$ and
$\delta_\rms(\bx,z)$ are the number density contrasts of centrals and
satellites at redshift $z$, respectively. Note that $\bar{n}_\rmc(z)$
and $\bar{n}_\rms(z)$ simply follow from Eq.~(\ref{avgngal}) by
replacing $\Phi(L|M)$ in Eq.~(\ref{avgnm}) by $\Phi_\rmc(L|M)$ and
$\Phi_\rms(L|M)$, respectively.

The detailed functional form that we adopt for $\Phi(L|M)$ is
discussed in \S\ref{sec:hos}. In this subsection we show how the CLF
enters in the computation of the (projected) galaxy-galaxy correlation
function, $w_\rmp(r_\rmp|L_1,L_2,z)$, and in the excess surface
density profile, $\Delta\Sigma(R|L_1,L_2,z)$.

Within the framework of the halo model, we can write
\begin{equation}
n_\rmc(\bx,z) = \sum_i \calN_{\rmh, i} \, \calN_{\rmc, i} \,
\delta^\rmD(\bx-\bx_i)\,,
\end{equation}  
where $\calN_{\rmc, i}$ is the number of central galaxies in the halo
whose center is in volume element $i$ (i.e., $\calN_{\rmc, i}$ is
either 0 or 1).  The Dirac delta function expresses the fact that a
central, by definition, resides at the center of a dark matter
halo. Similarly, for the satellite galaxies we can write
\begin{equation}\label{nsreal}
n_\rms(\bx,z) = \sum_i \calN_{\rmh, i} \calN_{\rms, i} u_\rms(\bx-\bx_i|M_i,z)\,,
\end{equation}  
where $\calN_{\rms, i}$ is a positive integer indicating the number of
satellite galaxies that reside in the halo whose center is in volume
element $i$, and $u_\rms(r|M,z)$ describes the {\it normalized} radial
distribution of satellite galaxies in an {\it average} halo of mass
$M$ at redshift $z$\footnote{Strictly speaking, by writing
$n_\rms(\bx,z)$ in terms of $u_\rms(r|M,z)$ we have already taken an
ensemble average over all possible spatial realizations of the
satellite galaxies in a halo of mass $M$ at redshift $z$. Hence, the
number density distribution of Eq.~(\ref{nsreal}) does not correspond
to a single realization, as it should. However, since we are only
concerned here with power-spectra, which are anyways based on ensemble
averaging, Eq.~(\ref{nsreal}) is adequate for what follows.}.

Using Eq.(\ref{ngalsplit}), the galaxy-galaxy power spectrum  can be
written as
\begin{eqnarray}\label{Pkgalsplit}
\lefteqn{P_{\rm gg}(k,z) = f^2_\rmc(z) P_{\rm cc}(k,z) + 
2 f_\rmc(z) f_\rms(z) P_{\rm cs}(k,z) } \nonumber \\
& & + f^2_\rms(z) P_{\rm ss}(k,z)\,,
\end{eqnarray}
while the galaxy-matter cross power spectrum is given by
\begin{equation}\label{Pkgaldmsplit}
P_{\rm gm}(k,z) = f_\rmc(z) P_{\rm cm}(k,z) + f_\rms(z) P_{\rm sm}(k,z)\,.
\end{equation}

Using the same methodology as in \S\ref{sec:halomodel} for the dark
matter, we split each of these five power-spectra into a 1-halo and a
2-halo term. The various 2-halo terms are given by
\begin{eqnarray}\label{P2hcc}
\lefteqn{P^{\rm 2h}_{\rmx\rmy}(k,z) =
\int \rmd M_1 \, \calH_\rmx(k|M_1,z) \, n(M_1,z) } \nonumber \\
& & \int \rmd M_2 \, \calH_\rmy(k|M_2,z) \, n(M_2,z) \,
Q(k|M_1,M_2,z)\,,
\end{eqnarray}
where `x' and `y' are either `c' (for central), `s' (for satellite),
or `m' (for matter), $Q(k|M_1,M_2,z)$ is given by Eq.~(\ref{QkM}), and we
have defined
\begin{equation}\label{calHm}
\calH_\rmm(k,M,z) = {M \over \bar{\rho}_{\rmm}} \,  \tilde{u}_\rmh(k|M,z)\,,
\end{equation}
\begin{equation}\label{calHc}
\calH_\rmc(k,M,z) = \calH_\rmc(M,z) = 
{\langle N_\rmc|M \rangle \over \bar{n}_{\rmc}(z)} \,,
\end{equation}
and
\begin{equation}\label{calHs}
\calH_\rms(k,M,z) = {\langle N_\rms|M \rangle \over \bar{n}_{\rms}(z)} \,  
\tilde{u}_\rms(k|M,z)\,.
\end{equation}
Here $\langle N_\rmc|M \rangle$ and $\langle N_\rms|M \rangle$ are the
average number of central and satellite galaxies in a halo of mass
$M$, which follow from Eq.~(\ref{avgnm}) upon replacing $\Phi(L|M)$ by
$\Phi_\rmc(L|M)$ and $\Phi_\rms(L|M)$, respectively. 

For the 1-halo terms, one obtains
\begin{equation}
P^{\rm 1h}_{\rm cc}(k,z) = {1 \over \bar{n}_\rmc(z)}\,,
\end{equation}
\begin{equation}
P^{\rm 1h}_{\rm cs}(k,z) = \int\rmd M \, 
\calH_\rmc(M,z) \, \calH_\rms(k,M,z) \, n(M,z) \,,
\end{equation}
and
\begin{equation}\label{p1hss}
P^{\rm 1h}_{\rm ss}(k,z) = \calA_\rmP 
\int\rmd M \, \calH^2_\rms(k,M,z) \, n(M,z) \,.
\end{equation}
Here we have assumed that the occupation numbers of centrals
and satellites are independent, so that $\langle N_\rmc N_\rms
|M\rangle = \langle N_\rmc|M \rangle \, \langle N_\rms|M \rangle$, and
we have introduced the parameter
\begin{equation}
\calA_\rmP \equiv {\langle N_\rms (N_\rms-1)|M\rangle \over 
\langle N_\rms|M \rangle^2}\,.
\end{equation}
If the occupation number of satellites follows a Poisson distribution,
i.e.,
\begin{equation}
P(N_\rms|M) = {\lambda^{N_\rms} \, \rme^{-\lambda} \over N_\rms !}\,,
\end{equation}
with $\lambda = \langle N_\rms|M\rangle$, then $\calA_\rmP = 1$, while
values of $\calA_\rmP$ larger (smaller) than unity indicate super- (sub-)
Poisson statistics. 

\subsection{The Projected Correlation Function and Excess Surface Density}
\label{sec:proj}

Once $P_{\rm gg}(k,z)$ and $P_{\rm gm}(k,z)$ have been determined, it
is fairly straightforward to compute the projected galaxy-galaxy
correlation function, $w_\rmp(r_\rmp,z)$, and the excess surface
density (ESD) profile, $\Delta\Sigma(R,z)$. We start by Fourier
transforming the power-spectra to obtain the two-point correlation
functions:
\begin{eqnarray}\label{xiFTfromPK}
\xi_{\rm xy}(r,z) & = & {1 \over (2\pi)^3} \int P_{\rm xy}(k,z) \
\rme^{+i \bk\cdot\bx} \rmd^3 \bk \nonumber \\
& = &
{1 \over 2 \pi^2} \int_0^{\infty} P_{\rm xy}(k,z) {\sin kr \over kr} \, k^2 \rmd k\,, 
\end{eqnarray}
where `x' and `y' are as defined above. 

As discussed above, the excess surface density profile 
\begin{equation}\label{shearb} 
\Delta\Sigma (R,z) = \overline{\Sigma}(<R,z) - \Sigma(R,z)\,,
\end{equation}
where $\overline{\Sigma}(<R,z) $ is given by Eq.~(\ref{averageSigma}).
The projected surface density, $\Sigma(R,z)$, is related to the
galaxy-matter cross correlation, $\xi_{\rm gm}(r,z)$, according
to
\begin{equation}\label{Sigma}
\Sigma(R,z) = \bar{\rho}_\rmm \int_{0}^{\omega_\rms} 
\left[1+\xi_{\rm gm}(r,z)\right] \, \rmd \omega \, ,
\end{equation}
where the integral is along the line of sight with $\omega$ the
comoving distance from the observer. The three-dimensional comoving
distance $r$ is related to $\omega$ through $r^2 = \omega_{\rm L}^2 +
\omega^2 - 2 \omega_{\rm L} \omega \cos \theta$. Here $\omega_{\rm L}$
is the comoving distance to the lens, and $\theta$ is the angular
separation between lens and source (see Fig.~1 in Cacciato \etal
2009). Since $\xi_{\rm gm}(r,z)$ goes to zero in the limit $r
\rightarrow \infty$, and since in practice $\theta$ is small, we can
approximate $\Sigma(R,z)$ using Eq.\,(\ref{Sigma_approx}), which
is the expression we adopt throughout.

The projected galaxy-galaxy correlation function is defined as
\begin{equation}\label{wpzspace}
w_\rmp(r_\rmp,z) = 2 \int_{0}^{r_{\rm max}} \xi_{\rm gg}(r_\rmp,r_{\pi},z) 
\,{\rm d}r_{\pi}\,.
\end{equation}
Here $r_\rmp$ is the projected separation between two galaxies,
$r_{\pi}$ is the redshift-space separation along the line-of-sight,
and $\xi_{\rm gg}(r_\rmp,r_{\pi},z)$ is the measured two-dimensional
correlation function, which is anisotropic due to the presence of
peculiar velocities. In the limit $r_{\rm max} \rightarrow \infty$,
the projected correlation function~(\ref{wpzspace}) is completely
independent of these peculiar velocities, simply because they have
been integrated out. In that case, $w_\rmp(r_\rmp)$ can be written
as a simple Abel transform of the real-space correlation function:
\begin{equation}\label{wprspace}
w_\rmp(r_\rmp,z) = 2 \int_{r_\rmp}^{\infty} \xi_{\rm gg}(r,z) \, 
{r \, {\rm d}r \over \sqrt{r^2 - r_\rmp^2}}\,,
\end{equation}
(Davis \& Peebles 1983).  However, since real data sets are always
limited in extent, in practice the projected correlation function
$w_\rmp(r_\rmp,z)$ is always obtained by integrating $\xi_{\rm
  gg}(r_\rmp,r_{\pi},z)$ out to some finite $r_{\rm max}$ rather than
to infinity. For example, Zehavi \etal (2011), whose data we use in
Paper~III, adopt $r_{\rm max} = 40 h^{-1} \Mpc$ or $60 h^{-1}\Mpc$,
depending on the luminosity sample used. This finite integration range
is often ignored in the modeling (e.g., Magliocchetti \& Porciani
2003; Collister \& Lahav 2005; Wake \etal 2008a,b) or is `accounted'
for by computing the model prediction for $w_\rmp(r_\rmp,z)$ using
Eq.~(\ref{wprspace}), but integrating from $r_\rmp$ out to $r_{\rm
  out} \equiv \sqrt{r_\rmp^2 + r^2_{\rm max}}$, where $r_{\rm max}$ is
the same value as used for the data (e.g., Zehavi \etal 2004, 2005,
2011; Abazajian \etal 2005; Tinker \etal 2005; Zheng \etal 2007, 2009;
Yoo \etal 2009). However, as we demonstrate in \S\ref{sec:zspace}
below, this introduces errors that can easily exceed 40 percent or
more on the largest scales probed by the data ($\sim 20h^{-1}\Mpc$;
see also Padmanabhan, White \& Eisenstein 2007; Norberg \etal 2009;
Baldauf \etal 2010). This is due to the fact that the peculiar
velocities on scales $r > r_{\rm max}$ cannot be ignored. In order to
take these residual redshift space distortions into account, we make
the assumption that the large scale peculiar velocities are completely
dominated by linear velocities (i.e., those that derive from linear
perturbation theory), and that the non-linear motions that give rise
to the Finger-of-God effect have been integrated out. In that case we
can correct Eq.~(\ref{wprspace}) for the fact that the projected
correlation function has been obtained using Eq.~(\ref{wpzspace}) with
a finite $r_{\rm max}$ as follows:
\begin{equation}\label{wpcorrect}
w_\rmp(r_\rmp,z) = 2 \, f_{\rm corr}(r_\rmp,z) \,
\int_{r_\rmp}^{r_{\rm out}} \xi_{\rm gg}(r,z) 
\, {r \, {\rm d}r \over \sqrt{r^2 - r^2_\rmp}}\,,
\end{equation}
where $f_{\rm corr}(r_\rmp,z)$ is the correction factor given by
\begin{equation}\label{fcorrect}
f_{\rm corr}(r_\rmp,z) = {
\int_0^{r_{\rm max}} \xi^{\rm lin}_{\rm gg}(r_\rmp,r_{\pi},z) \,\rmd r_{\pi}
\over \int_{r_\rmp}^{r_{\rm out}} \xi^{\rm lin}_{\rm gg}(r,z) 
{r \, {\rm d}r \over \sqrt{r^2 - r_\rmp^2}}}\,.
\end{equation}
Here $\xi^{\rm lin}_{\rm gg}(r,z)$ and $\xi^{\rm lin}_{\rm
  gg}(r_\rmp,r_{\pi},z)$ are the linear two-point correlation
functions of galaxies at redshift $z$ in real space and redshift
space, respectively.  For the former we may write
\begin{equation}\label{xiggrspace}
\xi^{\rm lin}_{\rm gg}(r,z) \equiv \bar{b}^2(z) \, \xi^{\rm lin}_{\rm mm}(r,z)\,,
\end{equation}
with $\xi^{\rm lin}_{\rm mm}(r,z)$ the two-point correlation function
of the initial matter field, linearly extrapolated to redshift $z$,
and
\begin{equation}\label{avbias}
\bar{b}(z) = {1 \over \bar{n}_\rmg(z)} \int \langle N_\rmg|M \rangle \,
b_\rmh(M,z) \, n(M,z) \, \rmd M\,,
\end{equation}
is the mean bias of the galaxies in consideration.  For the linear
galaxy correlation function in redshift space we can write
\begin{equation}\label{xiggzspace}
\xi^{\rm lin}_{\rm gg}(r_\rmp,r_{\pi},z) = 
\sum_{l=0}^2 \xi_{2l}(s,z)\,\calP_{2l}(\mu)
\end{equation}
(e.g., Kaiser 1987; Hamilton 1992). Here $s = \sqrt{r^2_\rmp +
  r^2_{\pi}}$ is the separation between the galaxies in redshift
space, $\mu = r_{\pi}/s$ is the cosine of the line-of-sight angle,
$\calP_l(x)$ is the $l^{\rm th}$ Legendre polynomial, and $\xi_0$,
$\xi_2$, and $\xi_4$ are given by
\begin{equation}\label{monopole}
\xi_0(r,z) = \left( 1 + {2 \over 3}\beta + {1 \over 5}\beta^2\right) \,
\xi^{\rm lin}_{\rm gg}(r,z)\,,
\end{equation}
\begin{equation}\label{quadrupole}
\xi_2(r,z) = \left( {4 \over 3}\beta + {4 \over 7}\beta^2\right) \,
\left[\xi^{\rm lin}_{\rm gg}(r,z) - 3 J_3(r,z)\right]\,,
\end{equation}
\begin{equation}\label{octopole}
\xi_4(r,z) = {8 \over 35}\beta^2 \, \left[\xi^{\rm lin}_{\rm gg}(r,z) + 
{15\over 2} J_3(r,z) - {35\over 2}J_5(r,z) \right]\,,
\end{equation}
where
\begin{equation}\label{Jintegral}
J_n(r,z) = {1 \over r^n} \int_0^r \xi^{\rm lin}_{\rm gg}(y,z) 
\, y^{n-1} \, \rmd y\,. 
\end{equation}
and
\begin{equation}\label{betapar}
\beta = \beta(z) = {1 \over \bar{b}(z)} 
\left({\rmd {\rm ln} D \over \rmd {\rm ln} a}\right)_z
\simeq {\Omega^{0.6}_\rmm(z) \over \bar{b}(z)}
\end{equation}
with $a = 1/(1+z)$ the scale factor and $D(z)$ the linear growth
rate. 

As we demonstrate in \S\ref{sec:zspace}, although this correction is
fairly accurate on large scales ($\gta 3 h^{-1}\Mpc)$, on smaller
scales it introduces an error of a few percent (see also Baldauf \etal
2010). Detailed tests with mocks indicate that this problem can be
avoided by simply replacing the linear galaxy-galaxy correlation
function in the Kaiser formalism with its non-linear analog; i.e., by
replacing in Eq.~(\ref{fcorrect}) and
Eqs.~(\ref{monopole})-(\ref{Jintegral}) each occurrence of $\xi_{\rm
  gg}^{\rm lin}(r,z)$ with $\xi_{\rm gg}(r,z)$ computed from
Eq.~(\ref{xiFTfromPK}) using the model outlined in
\S\ref{sec:xigg}. This is the method we will use throughout whenever
we compute $w_\rmp(r_\rmp,z)$ for comparison with data, always using
the same $r_{\rm max}$ as used for the data (see Paper~III) and with
$\bar{b}(z)$ computed from our CLF model using Eq.~(\ref{avbias}).
Note that with this modified version of the Kaiser formalism, the
denominator of $f_{\rm corr}$ in Eq.~(\ref{fcorrect}) is exactly equal
to the integral in Eq.~(\ref{wpcorrect}). Hence, there is no need to
compute the correction factor; rather, $w_\rmp(r_\rmp)$ can simply be
obtained directly using Eq.~(\ref{wpzspace}) with $\xi_{\rm
  gg}(r_\rmp,r_{\pi},z)$ given by
Eqs.~(\ref{xiggzspace})-(\ref{Jintegral}), but with $\xi^{\rm
  lin}_{\rm gg}(r,z)$ replaced by $\xi_{\rm gg}(r,z)$ (see
\S\ref{sec:zspace} for details).


\section{Model Ingredients}
\label{sec:ingredients}

The model described in the previous section requires a number of
ingredients, namely the halo mass function, $n(M,z)$, the halo bias
function, $b_\rmh(M,z)$, the radial bias function, $\zeta(r,z)$, the
linear and non-linear matter power spectra, $P^{\rm lin}_{\rm
  mm}(k,z)$ and $P_{\rm mm}(k,z)$, respectively, the (normalized) halo
density profile, $u_\rmh(r|M)$, the (normalized) radial number density
distribution of satellite galaxies, $u_\rms(r|M)$, and the halo
occupations statistics $\langle N_\rmc|M\rangle$ and $\langle
N_\rms|M\rangle$. We now discuss these ingredients in turn.

\subsection{Matter Power Spectra}
\label{sec:powerspec}

In our fiducial model, which includes a treatment of halo exclusion,
we require both the linear and the non-linear two-point correlation
functions of the matter, $\xi^{\rm lin}_{\rm mm}(r,z)$ and $\xi_{\rm
  mm}(r,z)$, which are the Fourier transform of the linear and
non-linear power-spectrum, $P^{\rm lin}_{\rm mm}(k,z)$ and $P_{\rm
  mm}(k,z)$, respectively. 

Throughout we compute $P_{\rm mm}(k,z)$ using the fitting formula of
Smith \etal (2003)\footnote{We use the small modification suggested on
  John Peacock's website http://www.roe.ac.uk/$\sim$jap/haloes/, although
  it has no significant impact on any of our results.} which is
modeled on the basis of the {\it linear} matter power spectrum,
\begin{equation}
P^{\rm lin}_{\rm mm}(k,z) \propto D^2(z) \, T^2(k) \, k^{n_\rms}\,.
\end{equation}
Here $n_\rms$ is the spectral index of the initial power spectrum,
$T(k)$ is the linear transfer function, and $D(z)$ is the linear
growth factor at redshift $z$, normalized to unity at $z=0$. We adopt
the linear transfer function of Eisenstein \& Hu (1998), which
properly accounts for the baryons, neglecting any contribution from
neutrinos and assuming a CMB temperature of 2.725K (Mather \etal
1999). The power spectrum is normalized such that the mass variance
\begin{equation}
\sigma^2(M) = {1 \over 2\pi^2} \int P^{\rm lin}_{\rm mm}(k,0) 
\widetilde{W}^2(kR) \, k^2 \rmd k\,, 
\end{equation}
is equal to $\sigma^2_8$ for $R=8 h^{-1} \Mpc$. Here
\begin{equation}
\widetilde{W}(kR) = {3 (\sin kR - kR \cos kR) \over (kR)^3}\,,
\end{equation}
is the Fourier transform of the spatial top-hat filter, and $M$ is
related to $R$ according to $M = 4\pi\bar{\rho}_{\rmm} R^3/3$.

\subsection{Halo Mass Function}
\label{sec:halomassfunc}

For the halo mass function, $n(M,z)$, which specifies the comoving
abundance of dark matter haloes of mass $M$ at redshift $z$, we use
the results of Tinker \etal (2008, 2010), who have shown that the
halo mass function is accurately described by
\begin{equation}\label{nmz}
n(M,z) = {\bar{\rho}_\rmm \over M^2} \, \nu \, f(\nu) \,
  {\rmd\ln\,\nu\over\rmd\ln M},
\end{equation}
where $\nu=\delta_{\rm sc}(z)/\sigma(M)$, with $\delta_{\rm sc}(z)$
the critical overdensity required for spherical collapse at $z$, and
\begin{equation}
f(\nu) = \eta_0 \, \left[ 1 + (\eta_1\nu)^{-2\eta_2} \right] \, 
\nu^{2\eta_3} \, \rme^{-\eta_4\nu^2/2}\,.
\end{equation}
For our definition of halo mass (see \S\ref{sec:halomodel}), Tinker
\etal (2010) find that $\eta_1 = 0.589 (1+z)^{0.20}$, $\eta_2 =-0.729
(1+z)^{-0.08}$, $\eta_3 =-0.243 (1+z)^{0.27}$, and $\eta_4 = 0.864
(1+z)^{-0.01}$, while $\eta_0 = \eta_0(z)$ is set by the normalization
condition
\begin{equation}
\int b_\rmh(\nu) \, f(\nu) \, \rmd\nu = 1 \,,
\end{equation}
with $b_\rmh(\nu)$ the halo bias function of Tinker \etal (2010),
specified in \S\ref{sec:halobiasfunc} below. This normalization
expresses that the distribution of matter is, by definition, unbiased
with respect to itself.

Throughout we adopt
\begin{equation}
\delta_{\rm sc}(z) = 0.15 \, (12\pi)^{2/3} \,
      {[\Omega_\rmm(z)]^{0.0055} \over D(z)}\,,
\end{equation}
which is a good numerical approximation to the critical threshold for
spherical collapse (Navarro, Frenk \& White 1997).

\subsection{Halo Bias Function}
\label{sec:halobiasfunc}

For the halo bias function we adopt the fitting function of Tinker
\etal (2010), which for our definition of halo mass, can be written as
\begin{eqnarray}\label{bhMz}
\lefteqn{b_\rmh(M,z) = b_\rmh(\nu) = } \nonumber \\
& & 1 - {\nu^{0.1325} \over \nu^{0.1325} + 1.0716} + 0.1830 \nu^{1.5}
+ 0.2652 \nu^{2.4}
\end{eqnarray}
where, as before, $\nu=\delta_{\rm sc}(z)/\sigma(M)$,

Although we believe the halo mass function and halo bias function
obtained by Tinker \etal (2008, 2010) to be the most accurate to date,
it is important to realize that they still can carry uncertainties
that can potentially impact cosmological results. It is unclear if
such uncertainties affect just the mass function normalization and
not its shape. We will carry out a proper investigation of this issue
in future work. Throughout this paper, however, we restrict ourselves to the
$n(M,z)$ and $b_\rmh(M,z)$ specified above.

\subsection{Radial Bias Function}
\label{sec:radbiasfunction}

An important ingredient of the halo model is the radial bias function,
$\zeta(r,z)$, which accounts for the fact that Eq.~(\ref{xihhdef})
becomes inaccurate in the quasi-linear regime, by making halo bias
scale dependent, i.e., it effectively describes the impact of the
non-zero higher-order bias factors in Eq.~(\ref{hmfield}).

Ideally, the radial dependence of the halo bias is to be computed from
first principles using, for example, (renormalized) perturbation
theory (e.g., Crocce \& Scoccimarro 2006; McDonald 2006,2007; Smith,
Scoccimarro \& Sheth 2007; Elia \etal 2011). However, it remains to be
seen whether these techniques can yield reliable results in the
quasi-linear regime of the 1-halo to 2-halo transition region, which
will probably require an impracticable large number of orders or loops
in the perturbation series. In the absence of such an analytical
solution we have to resort to empirical fitting functions calibrated
against numerical simulations. Throughout, we adopt the fitting
function of Tinker \etal (2005), given by
\begin{equation}\label{zetafit}
\zeta_0(r,z) = {[1 + 1.17 \, \xi_{\rm mm}(r,z)]^{1.49} 
\over [1 + 0.69 \, \xi_{\rm mm}(r,z)]^{2.09} }\,.
\end{equation}
The subscript 0 indicates that this fitting function was calibrated
using $N$-body simulations in which the haloes were identified using
the friends-of-friends (FOF) percolation algorithm (e.g., Davis \etal
1985), with a linking length of 0.2 times the mean inter-particle
separation. However, the halo mass function and halo bias function
used here are based on the spherical overdensity algorithm. As already
pointed out in Appendix~A of Tinker \etal (2012), because of these
different halo definitions, the fitting function~(\ref{zetafit}) is
likely to be inadequate on small scales, which we indeed find to be
the case (see \S\ref{sec:zetacal} below). After some trial and error,
while assuring an easy numerical implementation, we decided to adopt
the following, modified, radial bias function
\begin{equation}\label{zetamod}
\zeta(r,z) = \left\{ \begin{array}{ll}
    \zeta_0(r,z) & \mbox{if $r \geq r_{\psi}$} \\
    \zeta_0(r_{\psi},z) & \mbox{if $r < r_{\psi}$}
\end{array}\right.
\end{equation}
where the characteristic radius, $r_{\psi}$, is defined by
\begin{equation}\label{rpsidef}
\log\left[ \zeta_0(r_{\psi},z) \, \xi_{\rm mm}(r_{\psi},z) \right] = \psi
\end{equation}
where $\psi$ is a free parameter to be calibrated against numerical
simulations (see \S\ref{sec:zetacal}). Note that if
Eq.~(\ref{rpsidef}) has no solution, e.g., when $\psi \rightarrow
+\infty$, we set $r_{\psi}=0$, which corresponds to simply using the
fitting function~(\ref{zetafit}) without modification.

\subsection{Density Profile of Dark Matter Haloes}
\label{sec:densprof}

We assume that dark matter haloes are spheres whose normalized density
distribution is given by the NFW profile
\begin{equation}\label{NFW}
u_\rmh(r|M) = {\bar{\rho}_\rmm \over M} \,
{\delta_{200} \over (r/r_{*})(1+r/r_{*})^{2}}\,,
\end{equation}
(Navarro, Frenk \& White 1997), where $r_{*}$ is a characteristic
radius and $\delta_{200}$ is a dimensionless amplitude which can be
expressed in terms of the halo concentration parameter $c \equiv
r_{200}/r_{*}$ as
\begin{equation}\label{overdensity}
\delta_{200} = {200 \over 3} \, {c^{3} \over \ln(1+c) - c/(1+c)}\,.
\end{equation}
Numerical simulations show that $c$ is correlated with halo mass.
Throughout our work we use the concentration-mass relation of Macci\`o
\etal (2007), properly converted to our definition of halo mass.

The Fourier transform of the NFW profile, which features predominantly
in our model, is given by
\begin{eqnarray}\label{ukm}
\lefteqn{\tilde{u}_\rmh(k|M,z) = 
{3 \delta_{200} \over 200 c^3} \Bigl( 
\cos \mu \, [{\rm Ci}(\mu + \mu c) - {\rm Ci}(\mu)] \, + } \nonumber \\
& &  \sin \mu \, [{\rm Si}(\mu + \mu c) - {\rm Si}(\mu)] \, -
{\sin \mu c \over \mu + \mu c} \Bigr)\,,
\end{eqnarray}
where $\mu \equiv k r_{*}$, and ${\rm Si}(x)$ and ${\rm Ci}(x)$ are
the standard sine and cosine integrals, respectively. 

Note that this model for dark matter haloes is highly
oversimplified. In reality, haloes are triaxial, rather than
spherical, have scatter in the concentration-mass relation, have
substructure, and may have a density profile that differs
significantly from a NFW profile due to the action of baryons.  A
detailed discussion regarding the impact of these oversimplifications
on our results is presented in \S\ref{sec:assumptions}.

\subsection{Radial Number Density Distribution of Satellites}
\label{sec:nrsat}

Throughout, we assume that satellite galaxies follow a radial number
density distribution given by a generalized NFW profile (e.g., van den
Bosch \etal 2004):
\begin{equation}\label{generalisedNFW}
u_\rms(r|M) \propto \left({r\over \calR r_*} \right)^{-\gamma}
\left(1 + {r\over \calR r_*} \right)^{\alpha-3}\,,
\end{equation}
so that $u_\rms \propto r^{-\gamma}$ and $u_\rms \propto r^{-3}$ at
small and large radii, respectively. Here $\calR$ and $\gamma$ are two
free parameters, while the scale radius $r_*$ is the same as that for
the dark matter mass profile (Eq.~[\ref{NFW}]).  For our fiducial
model, we adopt $\calR = \gamma = 1$ so that $u_\rms(r|M) =
u_\rmh(r|M)$, i.e. satellites are unbiased with respect to the dark
matter. For consistency with our definition of halo mass, we only
adopt profile~(\ref{generalisedNFW}) out to $r_{200}$ (i.e., all
satellites have halo-centric radii $r < r_{200}$).

Observations of the number density distribution of satellite galaxies
in clusters and groups seem to suggest that $u_\rms(r|M)$ is in
reasonable agreement with an NFW profile, for which $\gamma=1$ (e.g.,
Beers \& Tonry 1986; Carlberg, Yee \& Ellingson 1997a; van der Marel
\etal 2000; Lin, Mohr \& Stanford 2004; van den Bosch \etal 2005a).
However, several studies have suggested that the satellite galaxies
are less centrally concentrated than the dark matter, corresponding to
$\calR > 1$ (e.g., Yang \etal 2005; Chen 2008; More \etal 2009a).  On
the other hand, in the case of very massive galaxies, in particular
luminous red galaxies, there are strong indications that they follow a
radial profile that is {\it more} centrally concentrated (i.e., $\calR
< 1$) than the dark matter (e.g., Masjedi \etal 2006; Watson \etal
2010, 2012; Tal, Wake \& van Dokkum 2012). In Paper~III we therefore
examine how the results depend on changes in $\calR$.

\subsection{Halo Occupation Statistics}
\label{sec:hos}

As specified in \S\ref{sec:xigg}, the halo occupation statistics
$\langle N_\rmc|M \rangle$ and $\langle N_\rms|M \rangle$, required to
describe the galaxy auto-correlation function and the galaxy-matter cross-correlation function, are obtained from the CLF, 
\begin{equation}
\Phi(L|M) = \Phi_\rmc(L|M) + \Phi_\rms(L|M)\, ,
\end{equation}

We use the CLF model presented in Cacciato \etal (2009), which is
motivated by the CLFs obtained by Yang, Mo \& van den Bosch (2008)
from a large galaxy group catalog (Yang \etal 2007) extracted from
the SDSS Data Release 4 (Adelman-McCarthy \etal 2006). In particular,
the CLF of central galaxies is modeled as a log-normal:
\begin{equation}\label{phi_c}
\Phi_\rmc(L|M) \,{\rmd}L = {\log\, e \over {\sqrt{2\pi} \, \sigma_\rmc}} 
{\rm exp}\left[- { {(\log L  -\log L_\rmc )^2 } \over 2\,\sigma_\rmc^2} \right]\,
{\rmd L \over L}\,,
\end{equation}
and the satellite term as a modified Schechter function:
\begin{equation}\label{phi_s}
\Phi_\rms(L|M)\,{\rmd}L = \phi^*_\rms \,
\left({L\over L^*_\rms}\right)^{\alpha_\rms + 1} \,
{\rm exp} \left[- \left ({L\over L^*_\rms}\right )^2 \right] {\rmd L \over L}\,,
\end{equation}
which decreases faster than a Schechter function at the bright end.
Note that $L_\rmc$, $\sigma_\rmc$, $\phi^*_\rms$, $\alpha_\rms$ and
$L^*_\rms$ are all functions of the halo mass $M$. 

Following Cacciato \etal (2009), and motivated by the results of Yang
\etal (2008) and More \etal (2009a, 2011, we assume that
$\sigma_\rmc$, which expresses the scatter in $\log L$ of central
galaxies at fixed halo mass, is a constant (i.e.  is independent of
halo mass and redshift).  In addition, for $L_\rmc$, which is defined
such that $\log L_\rmc$ is the expectation value for the (10-based)
logarithm of the luminosity of a central galaxy, i.e.
\begin{equation}
\log L_\rmc = \int_0^\infty \Phi_\rmc(L|M) \, \log L \, \rmd L\,,
\end{equation}
we adopt the following parameterization;
\begin{equation}\label{LcM}
L_\rmc(M) = L_0 {(M/M_1)^{\gamma_1} \over 
\left[1 + (M/M_1) \right]^{\gamma_1-\gamma_2}}\,.
\end{equation}
Hence, $L_\rmc \propto M^{\gamma_1}$ for $M \ll M_1$ and $L_c \propto
M^{\gamma_2}$ for  $M \gg  M_1$. Here $M_1$  is a  characteristic mass
scale, and $L_0 =  2^{\gamma_1-\gamma_2} L_c(M_1)$ is a normalization. 

For the satellite galaxies we adopt
\begin{equation}
L^*_\rms(M)  = 0.562 L_\rmc(M)\,,
\end{equation}
\begin{equation}\label{alpha}
\alpha_\rms(M) = \alpha_\rms
\end{equation}
and
\begin{equation}\label{phi}
\log[\phi^*_\rms(M)] = b_0 + b_1 (\log M_{12}) + b_2 (\log M_{12})^2\,,
\end{equation}
with $M_{12}=M/(10^{12} h^{-1}\Msun)$. Note that neither of these
functional forms has a physical motivation; they merely were found to
adequately describe the results obtained by Yang \etal (2008) from the
SDSS galaxy group catalog.
\begin{figure*}
\centerline{\psfig{figure=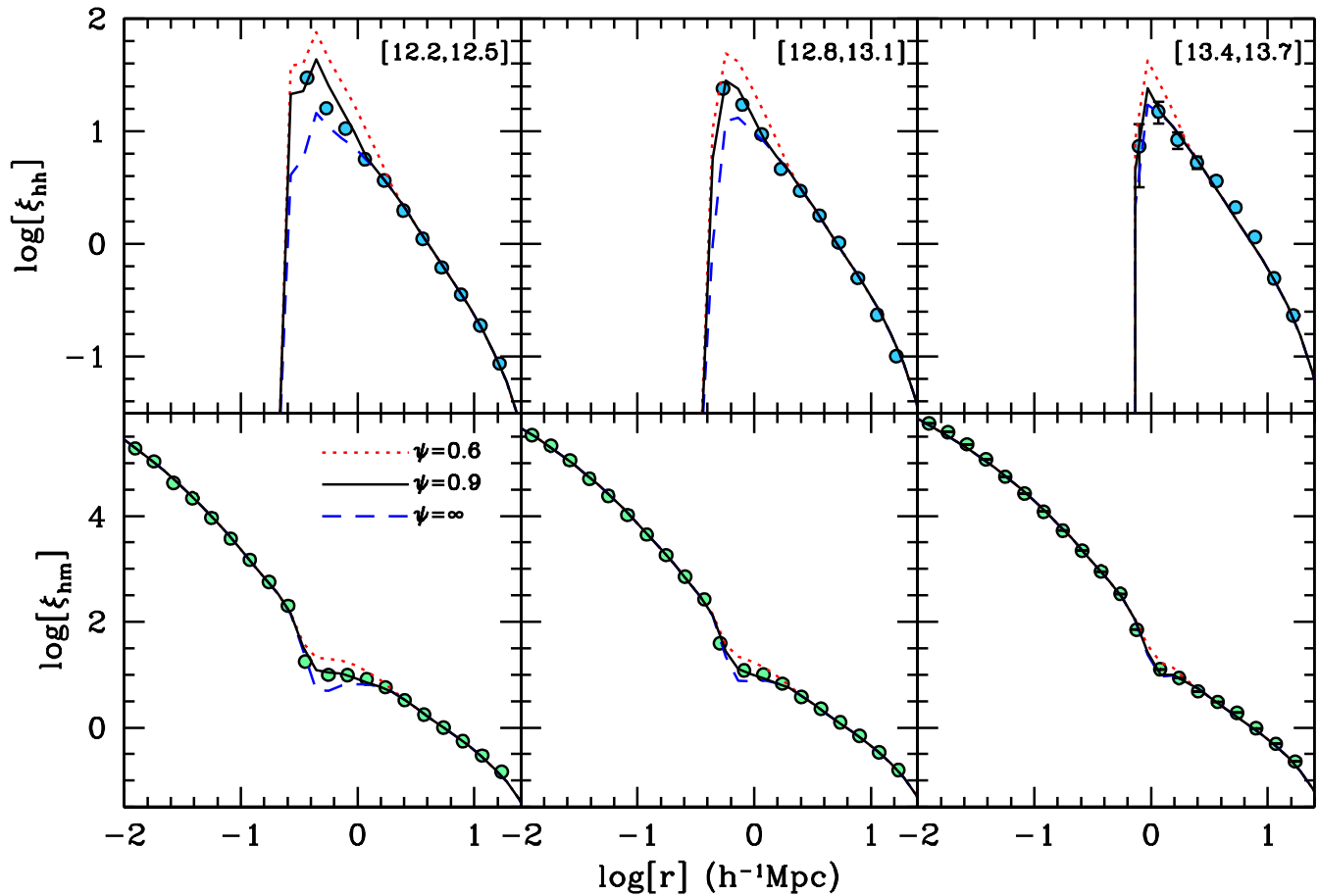,width=\hdsize}}
\caption{The halo-halo (top panels) and halo-matter (bottom panels)
  two-point correlation functions for haloes in three mass bins, as
  indicated in the top panels [values in square brackets in indicate
  $\log(M/(h^{-1}\Msun)$]. Colored symbols reflect the results
  obtained from the L250 simulation box. Errorbars (from Poisson
  statistics) are indicated, but since they are almost always smaller
  than the symbols, they can only be seen for 2 or 3 data points.  The
  various curves are analytical results for three different values of
  $\psi$, as indicated in the lower left-hand panel. Note that the
  model with $\psi=0.9$ accurately reproduces the sharp feature in
  $\xi_{\rm hm}(r)$, which reflects the 1-halo to 2-halo transition
  regime.}
\label{fig:xi_hh_hm}
\end{figure*}

Our parameterization of the CLF thus has a total of nine free
parameters
\begin{equation}\label{lamCLF}
\bolds\lambda_{\rm CLF} \equiv (\log M_1, \log L_{0}, \gamma_{1}, \gamma_{2}, 
\sigma_{\rm c}, \alpha_\rms, b_0, b_1, b_2)
\end{equation}

The final parameter used to describe the halo occupation statistics of
the galaxies is $\calA_\rmP$, defined in Eq.~(\ref{p1hss}). In our
fiducial model, adopted here, we will keep this parameter fixed at
$\calA_\rmP=1$, which corresponds to assuming that satellites follow
Poisson statistics.  As shown in Yang \etal (2008), this assumption
has strong support from galaxy group catalogs. Additional support
comes from numerical simulations which show that dark matter subhaloes
(which are believed to host satellite galaxies) also follow Poisson
statistics (Kravtsov \etal 2004). However, there are also some
indications that the occupation statistics of subhaloes and/or
satellite galaxies are actually slightly super-Poisson, i.e.,
$\calA_\rmP \gta 1$ (e.g., Porciani, Magliocchetti \& Norberg 2004;
Giocoli \etal 2010a; Busha \etal 2011; Boylan-Kolchin \etal
2010). Hence, in Paper~III we will also discuss models in which
$\calA_\rmP$ is taken to be a free parameter.
\begin{figure*}
\centerline{\psfig{figure=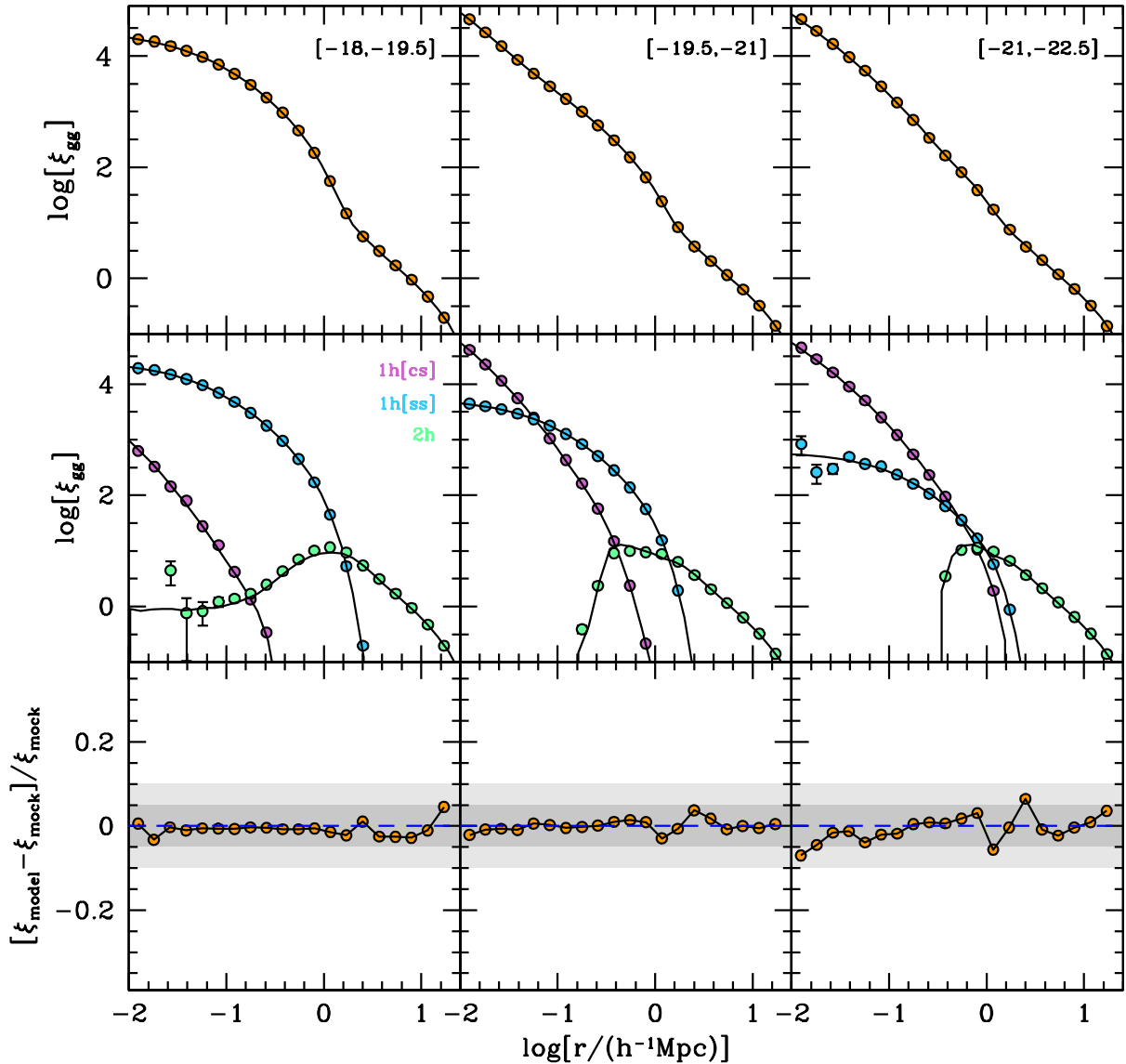,width=0.9\hdsize}}
\caption{Top panels show the galaxy-galaxy two-point correlation
  functions for three different magnitude bins, as indicated in the
  top panels [values in square brackets indicate $^{0.1}M_r - 5\log
  h$]. Colored symbols reflect the results obtained from the mock
  galaxy distribution in the L250 simulation box, while the solid line
  is the prediction of our analytical model. The middle panels show
  the contributions from the 1-halo central-satellite term (purple
  symbols, labeled `1h[cs]'), the 1-halo satellite-satellite term
  (blue symbols, labeled `1h[ss]'), and the 2-halo term (green
  symbols, labeled `2h'). Once again, the solid lines show the model
  predictions. As in Fig.~\ref{fig:xi_hh_hm}, errorbars reflecting
  Poisson statistics are indicated, but are almost always smaller than
  the symbols.  The bottom panels, show the fractional difference
  between model and mock for the total correlation functions shown in
  the top panels. The dark and light shaded areas indicate fractional
  errors of less than 5 and 10 percent, respectively. As is evident,
  the accuracy of our model is typically better than 5 percent, and
  always better than 10 percent.}
\label{fig:xigg_mock}
\end{figure*}


\section{Model Tests}
\label{sec:tests}

In this section we describe the construction of large mock galaxy
distributions, which we use to calibrate and test the real-space
galaxy-galaxy and galaxy-matter correlation functions computed using
the method outlined in \S\ref{sec:xigg}. In particular, we calibrate
the scale dependence of the halo bias and test the accuracy of our
halo-exclusion treatment, which we compare to some approximate methods
that do not account for halo exclusion but that are frequently used in
the literature. In addition, we also use these mock galaxy
distributions to test our correction for residual redshift space
distortions.
\begin{figure*}
\centerline{\psfig{figure=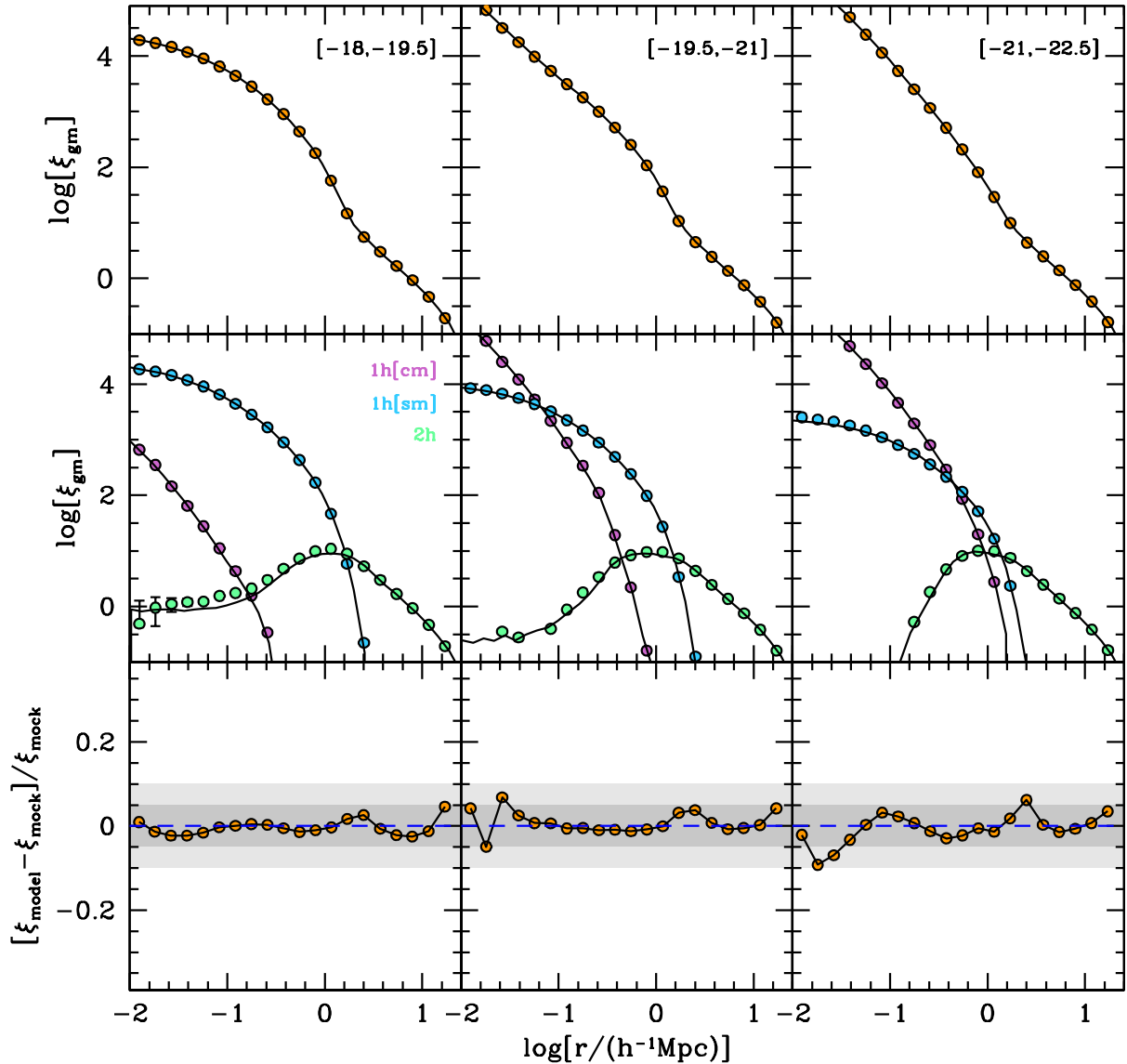,width=0.9\hdsize}}
\caption{Same as Fig.~\ref{fig:xigg_mock} but now for the
  galaxy-matter cross correlations. In the middle row of panels, the
  1-halo component is split in the central-matter (purple symbols,
  labeled `1h[cm]') and satellite-matter (blue symbols, labeled
  `1h[sm]') parts. Similar to the galaxy-galaxy correlation functions,
  the accuracy of our model is typically better than 5 percent, and
  always better than 10 percent.}
\label{fig:xigm_mock}
\end{figure*}

\subsection{Construction of Mock Galaxy Distributions}
\label{sec:mock}

For testing and calibrating the method described in \S\ref{sec:model}
we use two different $N$-body simulations that have been run using the
adaptive refinement technique (ART) of Kravtsov, Klypin \& Khokhlov
(1997). Both simulations have been used by Tinker \etal (2008, 2010)
in their studies of the halo mass function and halo bias function,
where they are called L250 and L1000W. We adopt the same nomenclature
in what follows.

Simulation L250 follows the evolution of 512$^3$ dark matter particles
in a cubic box of $250 h^{-1} \Mpc$ size in a flat $\Lambda$CDM
cosmology with matter density $\Omega_\rmm = 0.3$, baryon density
$\Omega_\rmb = 0.04$, Hubble parameter $h = 0.7$, spectral index
$n_\rms=1.0$, and a matter power spectrum normalization of $\sigma_8 =
0.9$.  Simulation L1000W follows the evolution of 1024$^3$ dark matter
particles in a $1 h^{-1} \Gpc$ size box in a flat $\Lambda$CDM
cosmology with matter density $\Omega_\rmm = 0.27$, baryon density
$\Omega_\rmb = 0.044$, Hubble parameter $h = 0.7$, spectral index
$n_\rms=0.95$, and a matter power spectrum normalization of $\sigma_8
= 0.79$. The particle masses are $m_\rmp = 9.69 \times 10^9 h^{-1}
\Msun$ and $m_\rmp = 6.98 \times 10^{10} h^{-1} \Msun$ for L250 and
L1000W, respectively.

For both simulations we use the halo catalogs at $z=0$, kindly
provided to us by Jeremy Tinker. These haloes are defined as spheres
with an overdensity of 200, which is identical to our definition of
halo mass (see \S\ref{sec:halomodel}).  More information about these
simulations and the identification of its dark matter haloes can be
found in Tinker \etal (2008).

In what follows we will use the L250 simulation box to calibrate and
test our galaxy-galaxy and galaxy-matter correlation functions, while
L1000W is used to test our correction for residual redshift space
distortions.  To this end, we construct mock galaxy distributions by
populating the dark matter haloes with model galaxies using the CLF. In
particular, we model the CLF using the parameterization described in
\S\ref{sec:hos} with the following parameters: $L_0 = 10^{9.9} h^{-2}
\Lsun$, $M_1 = 10^{10.9} h^{-1}\Msun$, $\sigma_\rmc = 0.16$,
$\gamma_1=5.0$, $\gamma_2=0.24$, $\alpha_s=-1.3$, $b_0=-1.2$,
$b_1=1.4$, and $b_2=-0.17$.  For each halo we first draw the
luminosity of its central galaxy from $\Phi_{\rm cen}(L|M)$, given by
Eq.~(\ref{phi_c}).  Next, we draw the number of satellite galaxies,
under the assumption that $P(N_{\rm sat}|M)$ follows a Poisson
distribution (i.e., $\calA_\rmP=1.0$) with mean
\begin{equation}\label{meanNsat}
\langle N_{\rm sat}|M\rangle = \int_{L_{\rm min}}^\infty \Phi_{\rm sat}(L|M) \, 
\rmd L\,,
\end{equation}
where we adopt a luminosity threshold, $L_{\rm min}$, corresponding to
$^{0.1}M_r - 5\log h = -18$ (here $^{0.1}M_r$ indicates the SDSS
$r$-band magnitude, $K$-corrected to $z=0.1$; see Blanton \etal
2003). For each of the $N_{\rm sat}$ satellites in the halo of
question we then draw a luminosity from the satellite CLF $\Phi_{\rm
  sat}(L|M)$, given by Eq.~(\ref{phi_s}).
\begin{figure*}
\centerline{\psfig{figure=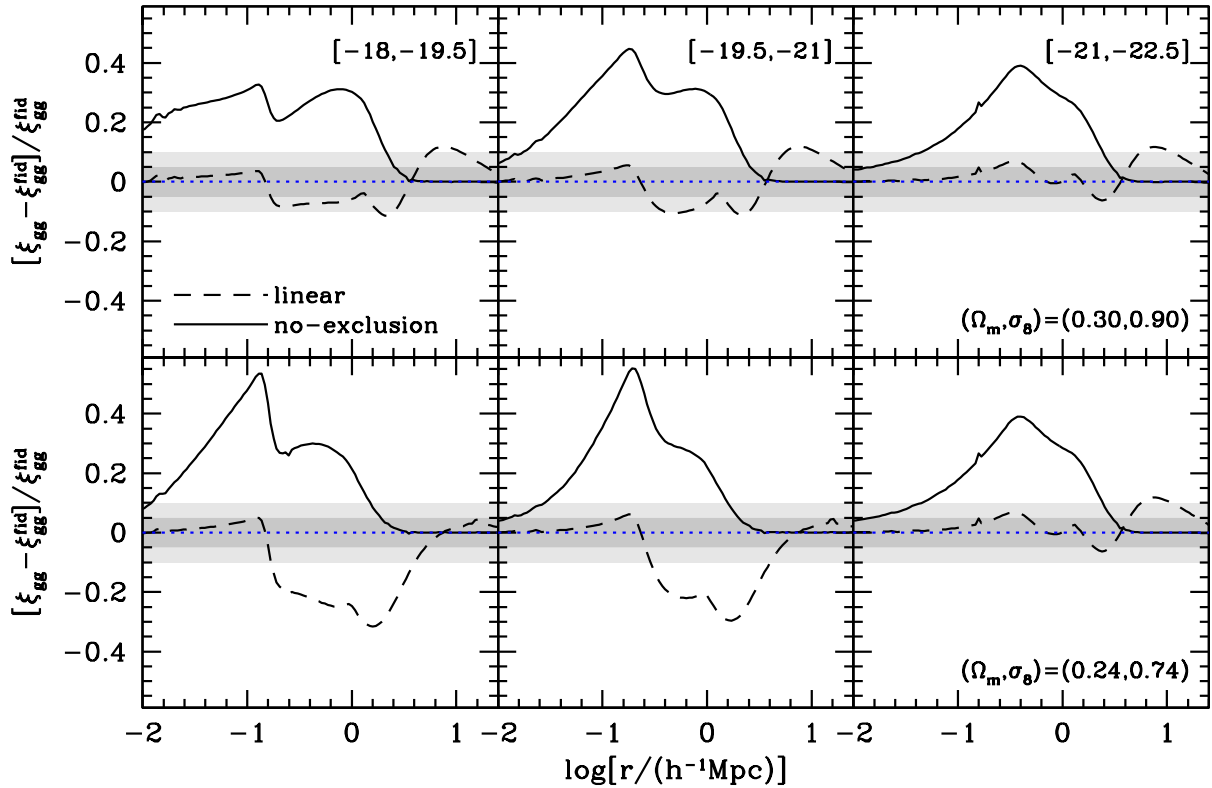,width=0.9\hdsize}}
\caption{The fractional errors of the approximate `no-exclusion'
model (solid lines) and `linear' model (dashed lines). Results
are shown for three magnitude bins, as indicated, and for two
different cosmologies+CLF. In the upper panels we use the same
cosmology and CLF as for the mocks in Figs.~\ref{fig:xi_hh_hm} -
\ref{fig:xigm_mock}; in the lower panels we use the WMAP3 cosmology
and the corresponding best-fit CLF model of Cacciato \etal (2009)
The dark and light shaded areas indicate fractional errors of less
than 5 and 10 percent, respectively.  Note that both the
`no-exclusion' model and the `linear' model have fractional errors
that can easily exceed 30-40 percent, which is inadequate for
precision cosmology.}
\label{fig:xi_comp}
\end{figure*}

Having assigned all mock galaxies their luminosities, the next step is
to assign them a position and velocity within their halo.  We assume
that the central galaxy resides at rest at the center of the halo,
while satellite galaxies follow a spherically symmetric number-density
distribution proportional to Eq.~(\ref{generalisedNFW}) with $\calR =
\gamma = 1$, i.e. we assume that satellite galaxies are unbiased with
respect to the dark matter. For the halo concentrations we adopt the
concentration-mass relation of Macci\`o \etal (2007), properly
converted to our definition of halo mass.  Finally, the peculiar
velocities of the satellite galaxies are assigned as follows. We
assume that satellite galaxies are in a steady-state equilibrium
within their dark matter potential well with an isotropic distribution
of velocities with respect to the halo center.  One dimensional
velocities are drawn from a Gaussian
\begin{equation}\label{vGauss}
f(v_j) = {1 \over \sqrt{2\pi}\,\sigma_{\rm sat}(r)} \,
\exp\left[-{v^2_j \over 2 \sigma^2_{\rm sat}(r)} \right]\,,
\end{equation}
with $v_j$ the velocity relative to that of the central galaxy along
axis $j$ and $\sigma_{\rm sat}(r)$ the local, one-dimensional velocity
dispersion obtained from solving the Jeans equation (see van den Bosch
\etal 2004; More \etal 2009b).

For reasons that will become clear below, in both simulation boxes we
only populate dark matter haloes with masses in the range $M_{\rm min}
\leq M \leq M_{\rm max}$, where $M_{\rm min} = 10^{12} h^{-1} \Msun$
and $10^{13} h^{-1} \Msun$ for L250 and L1000W, respectively, while
$M_{\rm max} = 10^{14.5}h^{-1} \Msun$ for both L250 and L1000W.

\subsection{Calibrating Scale Dependence of Halo Bias}
\label{sec:zetacal}

As discussed in \S\ref{sec:radbiasfunction}, fitting function
(\ref{zetafit}) for the radial bias is likely to be inaccurate on
small scales due to the fact that it was calibrated for a different
halo definition than the one used here. To investigate the magnitude
of this effect, and to test plausible corrections for it, we compare
our model predictions against the L250 simulation box.

We start by computing both the halo-halo auto-correlation function,
$\xi_{\rm hh}(r|M)$ and the halo-matter cross-correlation function,
$\xi_{\rm hm}(r|M)$, for a number of bins in halo mass.  We only
consider haloes in the mass range $10^{12} h^{-1}\Msun \leq M \leq
10^{14.5} h^{-1}\Msun$. The lower limit is needed to account for the
fact that the simulation has a finite mass resolution, while the upper
limit is adopted to be less sensitive to cosmic variance originating
from the relatively small volume of the simulation box. Over the mass
range $10^{12} h^{-1} \Msun \leq M \leq 10^{14.5} h^{-1} \Msun$ the
halo mass function is in excellent agreement with the fitting function
of Tinker \etal (2008), which is also the one used in our analytical
calculations.  Note that when cross-correlating the haloes with the
dark matter particles, we only consider the particles associated with
haloes in the mass range $10^{12} h^{-1}\Msun \leq M \leq 10^{14.5}
h^{-1}\Msun$: A large fraction of all dark matter particles in the
simulation box are not associated with any dark matter halo, but that
is simply a manifestation of the limited (mass and force) resolution
of the $N$-body simulation. In other words, the L250 simulation does
not properly resolve (non-linear) structure on a mass scale $M <
10^{12} h^{-1}\Msun$, and we therefore do not expect our model to
accurately reproduce the halo-matter cross correlation function of the
simulation if the cross correlation is with {\it all} dark matter. 

The resulting $\xi_{\rm hh}(r|M)$ and $\xi_{\rm hm}(r|M)$ are shown as
filled circles in the upper and lower panels of
Fig.~\ref{fig:xi_hh_hm}, respectively. The blue, dashed lines are our
model results, which are obtained using the same model as for the
galaxy-galaxy and galaxy-matter correlation functions described in
\S\ref{sec:xigg}, but by setting $\langle N_\rmc|M \rangle = 1$ if the
halo mass $M$ falls within the halo mass bin in consideration, and
$\langle N_\rmc|M \rangle = 0$ otherwise, plus $\langle N_\rms|M
\rangle = 0$ for all $M$. Note that all integrals over halo mass are
only integrated over the range $10^{12} h^{-1} \Msun \leq M \leq
10^{14.5} h^{-1} \Msun$.  Also, when Fourier transforming the
power-spectrum to obtain the correlation function, we adopt a lower
limit for the wavenumbers in order to account for the fact that the
simulation box has a finite size and periodic boundary conditions:
specifically, in Eq.~(\ref{xiFTfromPK}) we replace the lower limit of
the integration range by $k_{\rm min} = \sqrt{3} \times (2 \pi/L_{\rm
  box})$.  In this model we have set $\psi = +\infty$, which implies
that we have simply adopted the radial bias function of Tinker \etal
(2005) without any modification (i.e., $\zeta(r,z) = \zeta_0(r,z)$;
see \S\ref{sec:radbiasfunction}).

The model accurately fits the halo-matter cross correlation functions
on both small and large scales. The former indicates that our modeling
of the halo density profiles, $u(r|M)$, is accurate (i.e., we are not
making a significant error because we do not account for halo
triaxiality, halo substructure and scatter in halo concentration; see
\S\ref{sec:densprof}), while the good fit on large scales argues that
our treatment of halo bias is adequate. However, the model clearly
underpredicts $\xi_{\rm hm}(r)$ at the 1-halo to 2-halo transition
regime, which is especially conspicuous in the lower mass bin (lower
left-hand panel of Fig.~\ref{fig:xi_hh_hm}). The upper panels clearly
indicate that this is a reflection of the fact that the model
underpredicts the halo-halo correlation function on small scales
($\sim 1 h^{-1}\Mpc$; just before halo exclusion sets in).  The solid
and dotted lines are models in which we have used our modified version
of the radial bias function (Eq.~[\ref{zetamod}]) with $\psi = 0.9$
and $0.6$, respectively. The former provides the best-fit overall; it
somewhat overpredicts the halo-halo correlation function on small
scales in the lowest mass bin, but results in excellent fits to the
other correlation functions. The model with $\psi = 0.6$, on the
other hand, clearly overpredicts the small scale clustering of the
dark matter haloes for all mass bins. Detailed tests, including
additional halo mass bins and other functional forms for a modified
$\zeta(r,z)$, indicate that Eq.~(\ref{zetamod}) with $\psi = 0.9$
yield the best results, while still allowing for a sufficiently fast
numerical evaluation. We have also experimented with the modification
suggested by Tinker \etal (2012; see their Appendix~A), which is
identical to Eq.~(\ref{zetamod}), except that they adopt $r_{\psi} =
r_{200}(M_1) + r_{200}(M_2)$ rather than Eq.~(\ref{rpsidef}).  Not
only do we find this method to be less accurate, especially for the
lower mass bins, but the dependence of $r_{\psi}$ on halo mass also
makes the evaluation of $Q(k|M_1,M_2,z)$ more CPU intensive.

Note though, that there is no guarantee that $\psi = 0.9$ is also
the best-fit parameter for any cosmology other than the one considered
here. Hence, if we simply adopt $\psi = 0.9$ when trying to
constrain cosmological parameters, we might introduce an unwanted
systematic bias. Fortunately, as we demonstrate in Paper~II, $\psi$ is
only weakly degenerate with the cosmological parameters; most of its
degeneracy is with the parameters that describe the satellite
CLF. Hence, errors in $\psi$ may result in systematic errors in the
inferred satellite fractions, but will not significantly bias our
constraints on cosmological parameters. Nevertheless, in order to be
conservative, we will marginalize over uncertainties in $\psi$ when
fitting for cosmological parameters (see Paper~III).
\begin{figure*}
\centerline{\psfig{figure=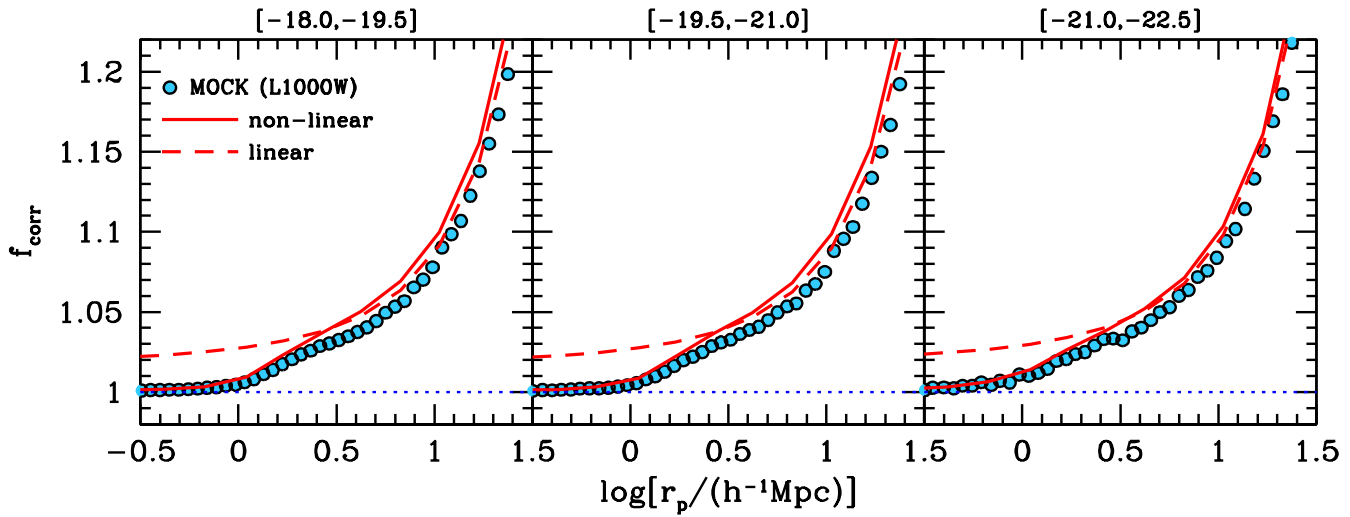,width=\hdsize}}
\caption{The correction factor, $f_{\rm corr}(r_\rmp)$, that describes
  the effect of residual redshift space distortions that arise from
  the use of a finite integration range when computing the projected
  correlation function, i.e., from Eq.~(\ref{wpzspace}) with a finite
  $r_{\rm max}$.  The shaded circles show the results obtained from
  the mock galaxy distribution in the L1000W simulation box with
  $r_{\rm max}=40 h^{-1}\Mpc$.  Results are shown for the same three
  magnitude bins as in Figs.~\ref{fig:xigg_mock} -
  \ref{fig:xi_comp}, as indicated.  Dashed and solid curves
  correspond to the $f_{\rm corr}(r_\rmp)$ obtained using the Kaiser
  formalism (see \S\ref{sec:proj}) with the linear and non-linear
  galaxy-galaxy correlation functions, respectively. The latter is in
  much better agreement with the mock results on small scales. See
  text for a detailed discussion.}
\label{fig:zspace_mock}
\end{figure*}

\subsection{Testing Halo Exclusion}
\label{sec:mgd}

Having calibrated the scale dependence of the halo bias, we now
proceed to test the accuracy of our model in calculating $\xi_{\rm
  gg}$ and $\xi_{\rm gm}$, focusing in particular on the accuracy of
our treatment of halo exclusion. Using the mock galaxy distribution
(hereafter MGD) of the L250 simulation box, we first compute the
real-space correlation function for three different luminosity
bins. The orange filled circles in the upper panels of
Fig.~\ref{fig:xigg_mock} show the results thus obtained. In the panels
in the middle row, we show the contribution to $\xi_{\rm gg}(r)$ from
the 2-halo term (green filled circles), the 1-halo central-satellite
term (purple filled circles) and the 1-halo satellite-satellite term
(blue filled circles). In the high-luminosity bin (right-hand panels),
the galaxy-galaxy correlation function is dominated by the 1-halo
central-satellite term on small scales ($r \lta 0.3 h^{-1} \Mpc$), and
by the 2-halo term on large scales ($r \gta 1.0 h^{-1} \Mpc$).  On
intermediate scales, the 1-halo satellite-satellite term dominates.
Note how this term becomes more and more dominant for less luminous
galaxies; in fact in the lowest luminosity bin considered here
(left-hand panels), the 1-halo satellite-satellite term completely
dominates the signal for $r \lta 1 h^{-1} \Mpc$. This reflects the
fact that the satellite fraction increases drastically from $f_{\rm
  sat} \simeq 0.136$ for the brightest bin, to $f_{\rm sat} \simeq
0.465$ for the intermediate luminosity bin, to $f_{\rm sat} \simeq
0.996$ for the faintest bin. Note, though, that these satellite
fractions are unrealistic due to the adopted cut-offs in halo mass at
$M = 10^{12} h^{-1} \Msun$ and $10^{14.5} h^{-1} \Msun$. For example,
for the CLF adopted here, virtually all central galaxies with $r$-band
magnitudes (K-corrected to $z=0.1$) in the range $-18 \geq
^{0.1}M_r - 5\log h \geq -19.5$ reside in haloes with $M < 10^{12}
h^{-1} \Msun$, which are not accounted for in our MGD; hence, almost
all mock galaxies in this magnitude range are satellites. For
comparison, if we were to integrate our CLF over the entire mass range
from $M=0$ to $M=\infty$, the corresponding satellite fractions, given
by
\begin{equation}
f_{\rm sat}(L_1,L_2) = {\int_{L_1}^{L_2} \rmd L \int_0^{\infty} \Phi_\rms(L|M) 
\, n(M) \, \rmd M \over \int_{L_1}^{L_2} \Phi(L) \rmd L}\,,
\end{equation}
are equal to $f_{\rm sat} = 0.334$, $0.253$, and $0.167$ from the
faintest to the brightest bin, respectively. Although the trends seen
in Fig.~\ref{fig:xigg_mock} are stronger than what is expected in
reality, we consider the fact that the dynamic range in $f_{\rm sat}$
covered is unrealistically large beneficial for the purpose of testing
the accuracy of our model.

The solid lines in the panels in the upper and middle rows of
Fig.~\ref{fig:xigg_mock} are the analytical results obtained using our
fiducial model with halo exclusion and with $\psi = 0.9$.  Here we
have adopted the same cosmology, redshift and CLF parameters as for
the MGD. Note that, once again, all integrals over halo mass are only
integrated over the range $10^{12} h^{-1} \Msun \leq M \leq 10^{14.5}
h^{-1} \Msun$, and we adopt $k_{\rm min} = \sqrt{3} \times (2
\pi/L_{\rm box})$ for the integration range in Eq.~(\ref{xiFTfromPK}).
Overall the agreement between our analytical prediction and the
results from the MGD is extremely good. As is evident from the panels
in the middle row, our treatment of halo exclusion nicely captures the
sudden decline of the 2-halo term on small scales. Although the
analytical 2-halo term becomes less accurate for $r \lta 0.5 h^{-1}
\Mpc$, mainly due to numerical issues, at these small scales the
1-halo term always dominates the total correlation function by at
least an order of magnitude. Hence, this inaccuracy is of little
practical concern. This is evident from the lower panels were we plot
the difference between the model prediction and the true correlation
function in the mock, normalized by the latter, as function of radius.
Over the entire range $0.01 h^{-1} \Mpc \leq r \lta 10 h^{-1}\Mpc$ the
model predictions agree with the mock results to an accuracy of a few
percent (typically $< 5\%$). At the 1-halo to 2-halo transition scale
($r \simeq 1 h^{-1} \Mpc$), which has been notoriously difficult to
model accurately, the errors are somewhat larger but always stay below
$10\%$.

Fig.~\ref{fig:xigm_mock} shows the same as Fig.~\ref{fig:xigg_mock},
but now for the galaxy-matter cross correlation, $\xi_{\rm gm}(r)$.
Similar trends are evident; the model's 2-halo term becomes less
accurate on small scales, but this has little to no impact on the
quality of the model as is evident from the lower panels. As for the
galaxy-galaxy correlation function, the model agrees with the
simulation results at the few percent level. In particular, it is
noteworthy that the model is accurate at better than 10 percent on
small scales.  This indicates that non-sphericity of haloes, scatter
in halo concentration, and halo substructure, all of which are ignored
in our model, do not have a large ($\gta 10$ percent) impact on the
results (see \S\ref{sec:assumptions} for a detailed discussion).

\subsection{Testing the Approximate Linear Model}
\label{sec:linmod}

As we have demonstrated above, our implementation of halo exclusion
and scale dependence of the bias are accurate at the few percent
level. However, the required computation of $Q(k|M_1,M_2,z)$, defined
in Eq.~(\ref{QkM}), is fairly CPU intensive.  The computation of
$w_\rmp(r_\rmp)$ and $\Delta\Sigma(R)$ for six luminosity bins (i.e.,
a single model; see paper~III) takes $\sim 20$ seconds on a single
(fast) processor. Consequently, the construction of an adequate Monte
Carlo Markov Chain (which has to be large given that our model has
anywhere from 14 to 19 free parameters, depending on the priors used)
takes several days to complete (on a single processor).  Although this
is not a major challenge in light of the fact that most desktop
computers nowadays have multiple processors, it nevertheless would be
hugely advantageous if a much faster, approximate method could be
found. In particular, the code can be made much faster if we were to
ignore halo exclusion and/or the scale dependence of the halo bias.

In this section we therefore investigate the pros (increase in speed)
and cons (decrease in accuracy) of two different simplifications of
our model. The first simplification is to ignore halo exclusion, i.e.,
we set $r_{\rm min} = 0$ in Eq.~(\ref{xihh}). In that case we have
that $\xi_{\rm hh}(r,z|M_1,M_2) = b_\rmh(M_1,z) \, b_\rmh(M_2,z) \,
\zeta(r,z) \, \xi_{\rm mm}(r,z)$, and the two-halo term of the power
spectrum~(\ref{P2hcc}) simplifies to
\begin{equation}\label{P2hrad}
P^{\rm 2h}_{\rm xy}(k,z) = \overline{b}_\rmx(k,z) \, \overline{b}_\rmy(k,z) \,
P_{\rm ne}(k,z) \,,
\end{equation}
where `x' and `y' are either `c' (for central), `s' (for satellite),
or `m' (for matter),
\begin{equation}\label{barbcz}
\bar{b}_\rmx(k,z) = \int\rmd M \, \calH_\rmx(k,M,z) \, n(M,z) \, 
b_\rmh(M,z) \,,
\end{equation}
with $\calH_\rmx(k,M,z)$ given by Eqs.~(\ref{calHm})--(\ref{calHs}), and
\begin{equation}\label{Prad}
P_{\rm ne}(k,z) = 4 \pi \int_0^{\infty} \zeta(r) \, \xi_{\rm mm}(r,z) \,
{\sin kr \over kr}\, r^2 \,\rmd r\,.
\end{equation}
This simplified model has the great advantage that it does not require
the tedious and CPU intensive evaluation of $Q(k|M_1,M_2,z)$, causing
a speed-up of a factor $\sim 10$, while still accounting for the scale
dependence of the halo bias. In what follows we shall refer to this
model as the `no-exclusion model'. The solid lines in
Fig.~\ref{fig:xi_comp} show the relative error in $\xi_{\rm gg}(r)$ of
the no-exclusion model with respect to our fiducial model with halo
exclusion. Results are shown for three magnitude bins, as indicated in
the top panels, and for two different cosmologies/CLFs. In the upper
panels we use the same cosmology and CLF as for the mocks described in
\S\ref{sec:mock}. In the lower panels we use the WMAP3 cosmology,
i.e., the cosmological parameters that best fit the three year data
release of the Wilkinson Microwave Anisotropy Probe (Spergel \etal
2007) and the best-fit CLF model for that cosmology obtained by
Cacciato \etal (2009). The main motivation for showing results for two
different cases is to emphasize that the fractional errors of the
no-exclusion model may vary quite significantly from one cosmology
and/or CLF to another.  Clearly the no-exclusion model in general
overpredicts the galaxy-galaxy correlation functions on small scales
($r \lta 2 h^{-1}\Mpc$) by 20 to 50 percent\footnote{The sharp
  features apparent around $0.3 h^{-1}\Mpc$ are not due to numerical
  noise, but are real manifestations of halo exclusion.}.  

At the risk of further deteriorating the accuracy of the model, we can
make additional simplifications by replacing $P_{\rm ne}(k,z)$ in
Eq.~(\ref{P2hrad}) by the linear matter power spectrum, $P^{\rm
  lin}_{\rm mm}(k,z)$. This results in the `linear' halo model, which
has been used previously by numerous authors (e.g., Ma \& Fry 2000;
Seljak 2000; Scoccimarro \etal 2001; Guzik \& Seljak 2002; Mandelbaum
\etal 2005; Seljak \etal 2005; see also Cooray \& Sheth 2002 and Mo
\etal 2010). This removes the need for the integration~(\ref{Prad})
and therefore further speeds up the computation, albeit at the cost of
ignoring the scale dependence of the halo bias. The dashed curves in
Fig.~\ref{fig:xi_comp} show how these `linear' galaxy-galaxy
correlation functions compare to the fiducial model with halo
exclusion and with scale dependence of halo bias.  Somewhat
surprisingly, for the cosmology+CLF shown in the upper panels, this
linear model performs significantly better than the no-exclusion
model, with errors that are always below 10 percent.  This indicates
that halo-exclusion and scale-dependence of halo bias have comparable
but opposite effects on small scales ($r \lta 1 h^{-1}\Mpc$), which
{\it may} roughly cancel each other. The lower panels, however, show
that this is not always the case, and that the linear model can
significantly underestimate the galaxy-galaxy correlation functions
(by as much as 30-40 percent) in the 1-halo to 2-halo transition
regime. In addition, the linear model typically overpredicts the
correlation power on large scales of $\sim 10 h^{-1} \Mpc$ by ~10
percent. This is a well known effect that has already been discussed
in numerous studies of the halo model (e.g., e.g., Ma \& Fry 2000;
Seljak 2000; Scoccimarro \etal 2001; Smith \etal 2003; Cole \etal
2005; Hayashi \& White 2008). Finally we note that similar tests for
the galaxy-matter cross correlation functions yield fractional errors
for the no-exclusion and linear models that are very similar as for
the galaxy-galaxy correlation functions shown in
Fig.~\ref{fig:xi_comp}.

Hence, despite the order of magnitude increase in computational speed,
we conclude that both the `no-exclusion' model and the `linear' model
suffer from systematic inaccuracies that can easily reach 30 to 40
percent, which we consider inadequate for the purpose of constraining
cosmological parameters. In Papers~II and III we therefore exclusively
use the much more accurate, but more CPU intensive, model described in
\S\ref{sec:model} above, which properly accounts for both halo
exclusion and scale dependence of the halo bias.

\subsection{Redshift Space Distortions}
\label{sec:zspace}

As discussed in \S\ref{sec:proj}, the projected correlation functions
used to constrain the models have been obtained using a finite range
of integration along the line-of-sight. Consequently, they suffer from
residual redshift space distortions (RRSDs) that need to be corrected
for.  In this section we investigate the magnitude of these RRSDs, as
well as the accuracy of our correction method, which is based on the
linear Kaiser formalism (Kaiser 1987). To that extent we use the mock
galaxy distribution (MGD) obtained from the L1000W simulation box, as
described in \S\ref{sec:mock}. We first use this MGD to compute the
projected correlation function, $w_\rmp(r_\rmp)$, for three luminosity
bins, by integrating the corresponding $\xi_{\rm gg}(r_\rmp,r_\pi)$
out to $r_{\rm max} = 40 h^{-1} \Mpc$\footnote{Here we have assumed
  that the plane-parallel approximation holds}. Note that this is the
same value of $r_{\rm max}$ as used by Zehavi \etal (2011) for
computing the projected correlation functions of faint galaxies in the
SDSS DR4. Next we compute the same $w_\rmp(r_\rmp)$, but this time we
set the peculiar velocities of all galaxies to zero, i.e., we simply
set $r_\pi = \sqrt{r^2 - r^2_\rmp}$, where $r$ is the real-space
separation between two galaxies. The ratio of these two `measurements'
of the projected correlation function, shown as filled circles in
Fig.~\ref{fig:zspace_mock}, indicates the error one makes in the
estimate of $w_\rmp(r_\rmp)$ when ignoring the RRSDs, i.e., when
computing $w_\rmp(r_\rmp)$ using
\begin{equation}\label{wprmax}
w_\rmp(r_\rmp) = 2 \int_{r_\rmp}^{r_{\rm out}} \xi_{\rm gg}(r) \, 
{r \, {\rm d}r \over \sqrt{r^2 - r_\rmp^2}}\,,
\end{equation}
with $r_{\rm out} = \sqrt{r_\rmp^2 + r^2_{\rm max}}$. As discussed in
\S\ref{sec:proj}, this is the standard method used by numerous
authors in the past. The MGD results in Fig.~\ref{fig:zspace_mock}
show that ignoring RRSDs causes an error in $w_\rmp(r_\rmp)$ that
exceeds 10 percent on scales $\gta 10 h^{-1} \Mpc$.  Note, though,
that in the MGD we only populated haloes in the mass range $10^{13}
h^{-1} \Msun \leq M \leq 10^{14.5} h^{-1} \Msun$. As we show below,
using the full mass range results in RRSDs that are even larger.
\begin{figure}
\centerline{\psfig{figure=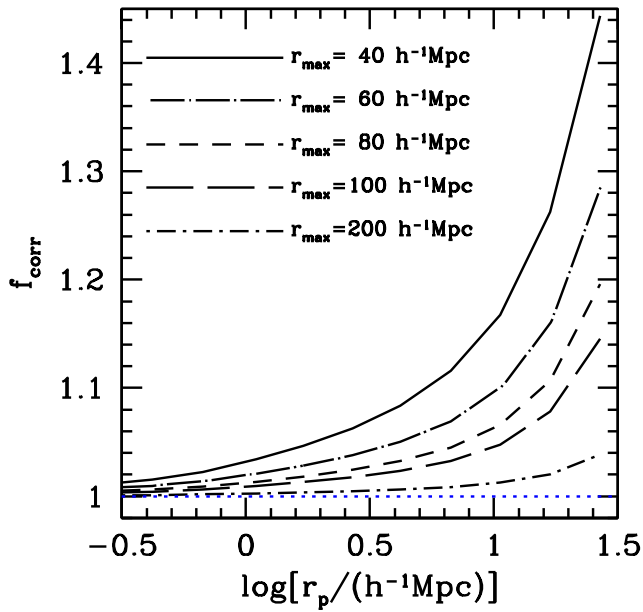,width=\hssize}}
\caption{The RRSD correction factor, $f_{\rm corr}(r_\rmp)$, for
  different values of the integration range $r_{\rm max}$, as
  indicated.  All these correction factors have been obtained for
  galaxies with $-21 \leq ^{0.1}M_r - 5\log h \leq -19.5$, assuming
  the same cosmology and CLF as for the L1000W mock (i.e., similar to
  the middle column in Fig.~\ref{fig:zspace_mock}). Note that $f_{\rm
    corr}$ for $r_{\rm max}=40 h^{-1}\Mpc$ is larger than in the case
  of Fig.~\ref{fig:zspace_mock}; this is due to the fact that here we
  integrate over all halo masses, whereas in
  Fig.~\ref{fig:zspace_mock} we only considered haloes with $10^{13}
  h^{-1}\Msun \leq M \leq 10^{14.5} h^{-1}\Msun$ in order to allow for
  a fair comparison with the mock results. Note also that even for
  $r_{\rm max}=200 h^{-1}\Mpc$ the correction factor exceeds 5 percent
  for $r_\rmp \gta 30 h^{-1}\Mpc$.}
\label{fig:zspace_rmax}
\end{figure}

The dashed line indicates the correction factor $f_{\rm corr}$ given
by Eq.~(\ref{fcorrect}). This correction factor is based on the Kaiser
formalism for the linear velocity field, and is computed using the
{\it linear} galaxy-galaxy correlation function given by
Eq.~(\ref{xiggrspace}). Note that the resulting $f_{\rm corr}$
provides a fairly accurate description of the RRSDs resulting from
using a finite $r_{\rm max}$, at least at large scales. However, on
small scales it clearly overpredicts $f_{\rm corr}$ by a few
percent. Hence, using this correction factor would overpredict
$w_\rmp(r_\rmp)$ by a similar amount on small scales.  

The solid line shows the correction factor obtained by simply
replacing $\xi_{\rm gg}^{\rm lin}(r)$ in Eq.~(\ref{fcorrect}) and
Eqs.~(\ref{xiggzspace})-(\ref{Jintegral}) by the non-linear version
$\xi_{\rm gg}(r)$. Although the Kaiser formalism is strictly only
valid in the linear regime, this simple modification works remarkably
well; the model now accurately reproduces the mock results on small
scales. On larger scales, the model somewhat overpredicts $f_{\rm
  corr}$ compared to the mock results. From the ratio between the two
we estimate that the final error we make on $w_\rmp(r_\rmp)$ from 
the imperfect correction for RRSDs is always less than 2 percent
over the scales of interest.

Finally, having demonstrated that $f_{\rm corr}(r_\rmp,z)$, obtained
using the non-linear galaxy-galaxy correlation function, provides an
accurate description of the RRSDs that arise from using a finite
integration range, we can use it to predict the magnitude of RRSDs for
different values of $r_{\rm max}$.  Fig.~\ref{fig:zspace_rmax} shows
$f_{\rm corr}(r_\rmp)$ for five different values of $r_{\rm max}$,
as indicated. Contrary to the results shown in
Fig.~\ref{fig:zspace_mock}, which only considered haloes in the mass
range $10^{13} h^{-1} \Msun \leq M \leq 10^{14.5} h^{-1} \Msun$ in
order to allow for direct comparison with the mock results, the
results in Fig.~\ref{fig:zspace_rmax} have been obtained by
integrating over all halo masses. Note that this results in $f_{\rm
  corr}$ values for $r_{\rm max} = 40 h^{-1}\Mpc$ that are
significantly larger than those in Fig.~\ref{fig:zspace_mock}.  In
particular, using $r_{\rm max} = 40 h^{-1}\Mpc$ without a correction
for RRSDs, underestimates $w_\rmp(r_\rmp)$ at $r_\rmp = 20 h^{-1}\Mpc$
by $\sim 35$ percent! Even when using $r_{\rm max} = 200 h^{-1} \Mpc$,
the RRSDs causes errors in the projected correlation function that
exceed $5$ percent for $r_\rmp \gta 30 h^{-1}\Mpc$. Clearly,
correcting for RRSDs is extremely important, especially when using
projected correlation functions to constrain cosmological parameters.
The modified Kaiser method presented here corrects for these RRSDs 
to an accuracy of better than 2 percent.


\section{Shapes, Alignment, Substructure and Contraction of
                Dark Haloes}
\label{sec:assumptions}

As discussed in \S\ref{sec:densprof}, our model assumes that dark
matter haloes are spheres with an NFW density profile. Clearly, this
is a highly oversimplified picture. In reality, dark matter haloes are
triaxial, have substructure, and have a density profile that may have
been modified due to the action of galaxy formation. In addition, our
model ignores the fact that there is significant scatter in the
relation between halo mass and halo concentration.  After discussing
how each of these effect impacts the accuracy of our oversimplified
model, we show how we can take these shortcomings into account by
marginalizing over the normalization of the concentration-mass
relation of dark matter haloes.

\subsection{Halo Shapes and Alignment}
\label{sec:haloshapes}

The assumption that dark matter haloes are spherical is inconsistent
with expectations based on numerical simulations (e.g., Jing \& Suto
2002; Bailin \& Steinmetz 2005; Allgood \etal 2006) and/or
non-spherical collapse conditions (e.g., Zel'dovich 1970; Icke 1973;
White \& Silk 1979). As shown by Yang \etal (2004), assuming that
haloes are spherical underestimates the correlation function obtained
if haloes are represented by FOF groups in numerical simulations by as
much as $\sim 20$ percent on small scales ($r \sim 0.1 h^{-1}
\Mpc$). A similar test was recently performed by van Daalen, Angulo \&
White (2011), who basically came to the same conclusion.  However,
these tests of the impact of halo triaxiality are not directly
applicable to our model. After all, our model uses halo mass functions
and halo bias functions in which haloes are specifically {\it defined}
as spherical volumes. Hence, a fair assessment of the impact of the
non-spherical symmetry of dark matter haloes on our results should
compare a correlation function in which it is assumed that all matter
within the spherical volume of the halo has spherical symmetry (i.e.,
our model assumption) to one in which the dark matter particles and
galaxies {\it within the same spherical volume} are given a more
realistic distribution that is not spherically symmetric. Note that
this is not the same as a comparison of spherical haloes to FOF
haloes, since the latter typically do not occupy a spherical
volume. As demonstrated by More \etal (2012, in preparation), this yields correlation
functions that only differ at the 5 to 10 percent level. Detailed
theoretical calculations by Smith \& Watts (2005) reach a similar
conclusion, that ignoring halo triaxiality only impacts the two-point
correlation functions at the level of $\sim 5$ percent. This is also
consistent with Li \etal (2009), who performed detailed tests that
showed that non-sphericity of dark matter haloes has only a small
effect of $\lta 5$ percent on the excess surface densities, and only
on the smallest scales probed by the data. Hence, we conclude that our
model assumption that haloes are spherical may underpredict both
$\xi_{\rm gg}(r)$ and $\xi_{\rm gm}(r)$ on small scales ($r < 1
h^{-1}\Mpc$), but by no more than $\sim 10$ percent.

However, the fact that haloes have triaxial, rather than spherical
shapes, also implies that another effect might in principle be
important, namely halo alignment. Such potential alignment between
haloes is not accounted for in our model, which therefore might cause
systematic errors in our two-point correlation functions. However,
Smith \& Watts (2005) have shown that a strict upper bound for the
effect of intrinsic alignment is a 10 percent effect on the two-point
correlation function (corresponding to a scenario with maximum
alignment). Van Daalen \etal (2011) have shown that realistic amounts
of alignment, as present in numerical simulations of structure
formation in a $\Lambda$CDM cosmology, has an effect on the
correlation functions that is not larger than $\sim 2$ percent. We
therefore conclude that potential halo alignment can be safely
ignored.

\subsection{Halo Concentrations}
\label{sec:haloconc}

As discussed in \S\ref{sec:densprof}, we assume that dark matter
haloes have NFW density profiles with a concentration-mass relation
given by Macci\`o \etal (2007), properly converted to our definition
of halo mass. This ignores, however, that there is a substantial
amount of scatter in the concentration-mass relation. In particular,
numerical simulations show that the concentrations, $c$, for haloes of
mass $M$ at redshift $z$ follow a log-normal distribution
\begin{equation}\label{pc}
P(c|M,z) \,\rmd c = {1 \over \sqrt{2\pi}\,\sigc} \, 
{\rm exp}\left[-{(\ln c - \ln\bar{c})^2 \over 2 \sigc^2}\right] \, 
{\rmd c \over c}\,,
\end{equation}
where $\bar{c} = \bar{c}(M,z)$ is the median halo concentration for a
halo of mass $M$ at redshift $z$, and $\sigc \simeq 0.3$ (Jing 2000;
Bullock \etal 2001; Wechsler \etal 2002; Sheth \& Tormen 2004; Macci\`o
\etal 2007). Because of this scatter, the proper
$\tilde{u}_\rmh(k|M,z)$ to use in the halo model is
\begin{equation}\label{ukprop}
\tilde{u}_\rmh(k|M,z) = \int \tilde{u}_\rmh(k|M,z,c) \, p(c|M,z) \, \rmd c
\end{equation}
(Giocoli \etal 2010b). However, in order to speed up the computations,
we ignore this scatter and simply use $\tilde{u}_\rmh(k|M,z) =
\tilde{u}_\rmh(k|\bar{c}(M,z))$ instead.
\begin{figure}
\centerline{\psfig{figure=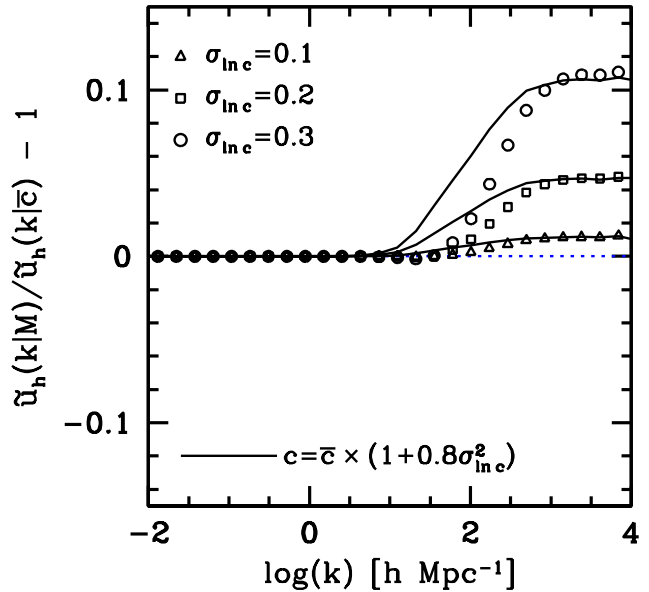,width=\hssize}}
\caption{The ratio $\tilde{u}_\rmh(k|M)/\tilde{u}_\rmh(k|\bar{c}) - 1$
  as function of the wavenumber $k$ for three different values of the
  scatter $\sigma_{{\rm ln}c}$ in $P(c|M)$, as indicated (open
  symbols). Here $\tilde{u}(k|M)$ is the Fourier Transform of the {\it
    average} normalized density profile of NFW haloes of mass $M$,
  properly accounting for the non-zero scatter in $P(c|M)$
  (Eq.~[\ref{ukprop}]), while $\tilde{u}_\rmh(k|\bar{c})$ is the
  normalized density profile for the median halo concentration,
  $\bar{c}$. Hence, this ratio indicates the error made in
  $\tilde{u}_\rmh(k|M)$ when ignoring the scatter in halo
  concentration. The solid lines show the same ratio, but this time
  $\tilde{u}(k|M)$ is computed under the assumption of zero scatter,
  and using a concentration parameter $c =
  \bar{c}\,(1+0.8\sigma^2_{{\rm ln}c})$. The reasonable agreement with
  the open symbols indicates that, to good approximation, one can
  mimic the effect of non-zero scatter in $P(c|M)$ by simply computing
  $\tilde{u}(k|M)$ for a halo concentration that is a factor
  $1+0.8\sigma^2_{{\rm ln}c}$ larger than the median concentration.}
\label{fig:cscatter}
\end{figure}

The impact of this oversimplification is shown in
Fig.~\ref{fig:cscatter}, where the symbols show $\tilde{u}_\rmh(k|M,z)
/ \tilde{u}_\rmh(k|\bar{c}(M,z)) - 1$, with $\tilde{u}_\rmh(k|M,z)$
given by Eq.~(\ref{ukprop}).  Results are shown for three different
values of $\sigc$, as indicated, and are obtained using $M = 10^{12}
h^{-1}\Msun$ and $\bar{c} = 10$. Taking the scatter in halo
concentration into account boosts $\tilde{u}_\rmh(k)$ on small scales
($k \gta 10 h \Mpc^{-1}$) by an amount that increases with $\sigc$
(see also Cooray \& Hu 2001 and Giocoli \etal 2010b). For $\sigc =
0.3$ this boost is of the order of 10 percent.  The solid lines in
Fig.~\ref{fig:cscatter} show $\tilde{u}_\rmh(k|c) /
\tilde{u}_\rmh(k|\bar{c}) - 1$, where $c = \bar{c} \, (1 + 0.8
\sigc^2)$. Although certainly not a perfect fit, this simple relation
gives a reasonable description of the impact of ignoring the scatter
in $p(c|M,z)$. It shows that for $\sigc = 0.3$, the error made
ignoring this scatter is similar to the error made if $\bar{c}(M,z)$
is underestimated by a factor $1 + 0.8 \sigc^2 \simeq 1.07$.  This is
comparable to the differences in the $\bar{c}(M,z)$ relation obtained
by different authors (e.g., Eke, Navarro \& Steinmetz 2001; Bullock
\etal 2001; Macci\'o \etal 2007; Zhao \etal 2009). Hence, it is at
least as important to obtain a more reliable calibration of the median
of $p(c|M,z)$ than to take account of its scatter. As we discuss in
\S\ref{sec:marginalization} below, because of these uncertainties, and
because of other oversimplifications of our model, we will marginalize
over the normalization of the concentration-mass relation,
$\bar{c}(M,z)$, when constraining cosmological parameters (see
Paper~III). The results shown here indicate that such a
marginalization also captures the inaccuracies arising from the fact
that we ignore the scatter in $p(c|M)$.

\subsection{Halo Substructure}
\label{sec:halosub}

Another oversimplification of our model is that we assume that dark
matter haloes have a smooth density distribution. However numerical
simulations of hierarchical structure formation have shown that haloes
are not smooth, but have a significant population of dark matter
subhaloes (e.g., Moore \etal 1998; Springel \etal 2001). Approximately
10 percent of the mass of a dark matter halo is associated with these
subclumps, with a weak dependence on halo mass and cosmology (e.g.,
Gao \etal 2004; van den Bosch \etal 2005b; Giocoli \etal 2008, 2010a).
Since these subhaloes are believed to host satellite galaxies, they
will impact the galaxy-matter cross correlation function on small
scales.  Although formalisms to include dark matter substructure in
the halo model have been developed (e.g., Sheth \& Jain 2003; Giocoli
\etal 2010b), the implementation is numerically cumbersome in that it
adds a number of integrations, causing a very significant increase in
the computation time per model. In addition, the model still involves
a number of uncertainties, such as the density profiles of dark matter
subhaloes.

Fortunately, as shown by Mandelbaum \etal (2005), Yoo \etal (2006) and
Li \etal (2009), the impact of substructure is negligible on the
radial scales of interest, i.e., on the scales for which we currently
have data on $\Delta\Sigma(R)$ available ($R \gta 0.05
h^{-1}\Mpc$). Hence, we conclude that we do not make significant
errors by ignoring dark matter substructure.

\subsection{The Impact of Baryons}
\label{sec:ac}

Although numerical simulations of structure formation have established
that dark matter haloes follow a universal profile that is accurately
described by the NFW profile (Eq.~[\ref{NFW}]), this ignores the
impact of baryons.  During the process of galaxy formation, baryons
collect at the center of the halo potential well and may subsequently
be expelled due to feedback processes. Because of the gravitational
interaction between baryons and dark matter, the dark matter halo will
respond to this galaxy formation process.

It is often assumed that the impact of baryons is to cause (adiabatic)
contraction of the dark matter haloes (e.g., Blumenthal \etal 1986;
Gnedin \etal 2004; Abadi \etal 2010; see also Schulz, Mandelbaum \&
Padmanabhan 2010; More \etal 2012b for observational support).
However, it is also possible
for haloes to expand in response to galaxy formation; rapid mass-loss
from the galaxy due to (repetitive) feedback from supernovae and/or
AGN (e.g., Pontzen \& Governato 2012), dynamical friction operating on
baryonic clumps (e.g., El-Zant, Shlosman \& Hoffman 2001; Mo \& Mao
2004), and galactic bars (e.g., Weinberg \& Katz 2002) all may cause
dark matter haloes to become less centrally concentrated than their
`pristine' (i.e., without galaxy formation) counterparts.

Interestingly, both galaxy rotation curves and galaxy scaling
relations suggest that dark matter haloes are less centrally
concentrated than what is expected in the absence of baryonic
processes in a CDM dominated universe (e.g., Swaters \etal 2003; de
Blok \etal 2008; Dutton \etal 2007, 2011; Trujillo-Gomez \etal
2011). Although this may suggest that galaxy formation indeed results
in a net halo expansion, it may also indicate that dark matter is not
dark, but warm (e.g., Sommer-Larsen \& Dolgov 2000) or
self-interacting (e.g., Spergel \& Steinhardt 2000).

We conclude that the detailed density profiles of dark matter haloes
carry a significant uncertainty, which needs to be accounted for.

\subsection{Marginalization}
\label{sec:marginalization}

All the effects discussed above, regarding halo shape, scatter in halo
concentrations, halo substructure, and halo contraction/expansion,
impact the 1-halo terms of the correlation functions by either
boosting or suppressing power on small scales.  What is ultimately of
importance for the accuracy of our models is the {\it combined} impact
of all these effects.
\begin{figure}
\centerline{\psfig{figure=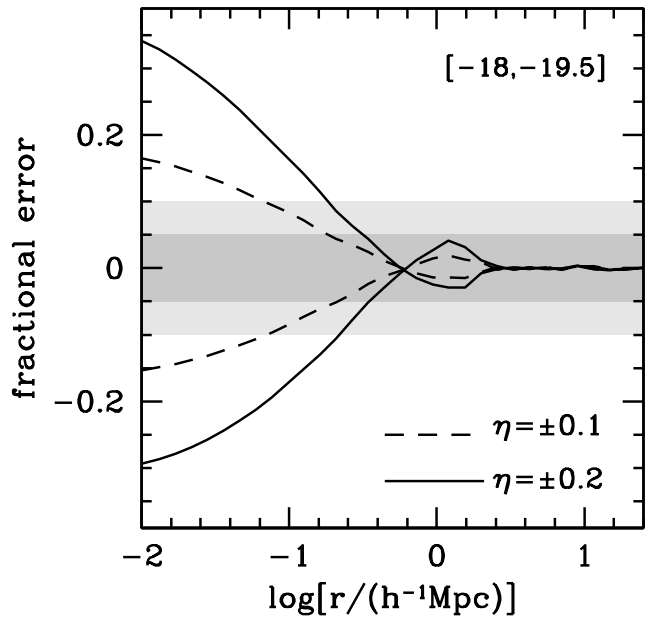,width=\hssize}}
\caption{The impact on the galaxy-matter cross correlation
  function, $\xi_{\rm gm}(r)$ of multiplying the normalization of the
  concentration-mass relation, $c(M)$, of dark matter haloes by a
  factor $(1+\eta)$, where $\eta = \pm 0.1$ (dashed lines) or
  $\eta = \pm 0.2$ (solid lines). Here we have, once again, adopted
  the same cosmology and CLF as for the mocks described in
  \S\ref{sec:mock}.}
\label{fig:cnorm}
\end{figure}

The combined impact of all effects except for that of halo
contraction/expansion can be gauged from the lower panels of
Fig.~\ref{fig:xigm_mock}, which show that our model is consistent with
the simulation results, in which the haloes have realistic, triaxial
density distributions, have substructure, and have non-zero scatter in
the concentration-mass relation, to better than 10 percent. This test
therefore confirms that our oversimplifications are accurate at the 10
percent level.  

We caution, though, that this test does not account for possible halo
contraction/expansion due to baryons, whose impact is difficult to
gauge in the absence of a more detailed understanding of galaxy
formation.  Hence, when constraining cosmological parameters (see
Paper~III), we will take all these oversimplifications regarding the
density distributions of dark matter haloes into account by
marginalizing over the normalization of the concentration-mass
relation, $\bar{c}(M,z)$. In particular, we introduce the parameter
$\eta$, so that the concentration for a halo of mass $M$ is given by
$(1+\eta) \times \bar{c}(M,z)$, where $\bar{c}(M,z)$ is the average
concentration-mass relation of Macci\`o \etal (2007), properly
converted to our definition of halo mass. As a prior we assume that
the probability distribution function (PDF) for $\eta$ is given by
\begin{equation}\label{fcprior}
P(\eta) = {1 \over \sqrt{2\pi}\sigma_{\eta}} \exp\left(-{\eta^2\over 
2 \, \sigma^2_{\eta}}\right)
\end{equation}
where we adopt $\sigma_{\eta} = 0.1$. Fig.~\ref{fig:cnorm} shows the
impact of $\eta$ on the galaxy-matter cross-correlation function for
galaxies with magnitudes in the range $-18 \geq {^{0.1}M}_r - 5\log h
\geq -19.5$ (results for other magnitude bins are very similar).  The
dashed and solid lines show the fractional changes in $\xi_{\rm
  gm}(r)$ for $\eta = \pm 0.1$ and $\pm 0.2$, respectively, which
correspond to the 68 and 95 percent confidence intervals of the prior
PDF. Note how $\eta = \pm 0.2$ modifies the one-halo term of
$\xi_{\rm gm}(r)$ by more than 20 percent on small scales ($r < 0.1
h^{-1}\Mpc$), which we argue is more than adequate to capture the
inaccuracies in our model that arise from the various
oversimplifications discussed above (see Paper~III for more details,
and for a discussion of the posterior distribution of $\eta$ and its
implications).


\section{Conclusions}
\label{sec:concl}

Galaxies are abundant and visible to high redshifts, making them, in
principle, excellent tracers of the mass distribution in the Universe
over cosmological scales. The problem, however, is that galaxies are
biased tracers, and that this bias is a complicated function of scale,
luminosity, morphological type, etc. It is an imprint of the poorly
understood physics related to galaxy formation.  On sufficiently large
scales, galaxy bias is expected to be scale-independent with a value
that is known to depend on a variety of galaxy properties such as
luminosity and color (e.g., Norberg \etal 2001, 2002; Zehavi \etal
2005, 2011; Wang \etal 2007).  On small, (quasi) non-linear scales ($r
\lta 3 h^{-1} \Mpc$), galaxy bias becomes strongly scale-dependent
(e.g., Cacciato \etal 2012a), making it extremely difficult to infer
any constraints on cosmology, without having a proper, detailed method
of either measuring this bias or marginalizing over it.  For this
reason, almost all studies to date that used the distribution of
galaxies in order to constrain cosmological parameters have focused on
large, linear scales, and treated galaxy bias as a `nuisance
parameter' that needs to be marginalized over.

In this paper, the first in a series, we have presented a new method,
similar to that of Yoo \etal (2006) and Leauthaud \etal (2011),
that can simultaneously solve for cosmology and galaxy bias on small,
non-linear scales. The method uses the halo model to analytically
describe the (non-linear) matter distribution, and the
conditional luminosity function (CLF) to specify the halo occupation
statistics. For a given choice of cosmological parameters, which
determine the halo mass function, the halo bias function, and the
(non-linear) matter power spectrum, this model can be used to predict
the galaxy luminosity function, the two-point correlation functions of
galaxies as function of both scale and luminosity, and the
galaxy-galaxy lensing signal, again as function of both scale and
luminosity.  These are all observables that have been measured at
unprecedented accuracies from the Sloan Digital Sky Survey, and can
therefore be used to constrain cosmological parameters.

In this paper we presented, in detail, our analytical framework, which
is characterized by
\begin{itemize}
\item a treatment for scale dependence of halo bias on small scales,
  using a modified version of the empirical fitting function of Tinker
  \etal (2005).
\item a proper treatment for halo exclusion, similar to that of Smith
  \etal (2007), which is correct under the assumption that dark matter
  haloes are spherical.
\item a correction for residual redshift space distortions (RRSDs)
  using a slightly modified version of the linear Kaiser formalism.
\end{itemize}
We have tested the accuracy of our analytical model using detailed
mock galaxy distributions, constructed using high-resolution numerical
$N$-body simulations.  We have shown that our analytical model is
accurate to better than 10 percent (in most cases better than 5
percent), in reproducing the 3-dimensional galaxy-galaxy correlation
and the galaxy matter correlation in the mock galaxy distributions
over a wide range of scales ($0.03
h^{-1} \Mpc \lta r \lta 30 h^{-1} \Mpc$).  In order to reach this
level of accuracy we had to introduce, and tune, one free parameter
that describes a modification of the empirical fitting function of
Tinker \etal (2005) for the radial halo bias dependence. This
modification is required because this fitting function is only valid
for a particular definition of halo mass that is different than the
one adopted here (see also Tinker \etal 2012).  When fitting the data
in order to constrain cosmological constraints, we will marginalize
over uncertainties in this free parameter (see Papers~II and~III). We
have demonstrated that ignoring halo exclusion and/or the scale
dependence of the halo bias results in errors in $\xi_{\rm gg}(r)$ and
$\xi_{\rm gm}(r)$ in the 1-halo to 2-halo transition regime ($r \sim 1
h^{-1}\Mpc$) that can easily be as large as 40 percent. The correction
for RRSDs is necessary because projected correlation functions are
always obtained by integrating along the line-of-sight out to a finite
radius (typically $r_{\rm max} \sim 40 - 80 h^{-1}\Mpc$) rather than
out to infinity. In agreement with the results of Norberg \etal
(2009), we show that not taking these RRSDs into account results in
systematic errors that can easily exceed 20 percent on large scales
($r_\rmp \gta 10 h^{-1}\Mpc$), which can cause systematic errors in
the inferred galaxy bias (see More 2011).  As we demonstrate in
Paper~III, when unaccounted for these RRSDs can also result in
significant systematic errors in the inferred cosmological
parameters. Fortunately, as we have demonstrated, it is fairly
straightforward to correct for these RRSDs, to an accuracy better than
$\sim 2$ percent, using a mildly modified version of the linear Kaiser
formalism (Kaiser 1987).

Finally, the good accuracy of our analytical model on small scales for
the galaxy-matter and halo-matter cross correlation functions (better
than 10 percent) indicates that ignoring halo triaxiality, halo
substructure, and scatter in the halo concentration-mass relation does
not have a large impact, contrary to recent claims by van Daalen \etal
(2011) who argue that halo triaxiality alone may cause inaccuracies as
large as 20 percent. We argue that this apparent discrepancy mainly
owes to different definitions of dark matter haloes (see discussion in
\S~\ref{sec:haloshapes}). Nevertheless, we have shown that, in order
to be conservative, one can take these inaccuracies that arise from
oversimplifications of the halo mass distributions into account by
marginalizing over uncertainties in the normalization of the
concentration-mass relation of dark matter haloes.

As indicated above, this is the first paper in a series.  In Paper~II
(More \etal 2012a), we perform a Fisher matrix analysis to (i)
investigate the strength of each of the datasets
(luminosity function, projected correlation functions, and excess
surface densities), (ii) identify various degeneracies between our
model parameters, and (iii) forecast the accuracy with which various
cosmological parameters and CLF parameters can be constrained with
current data. In Paper~III (Cacciato \etal 2012b) we apply our method
to data from the Sloan Digital Sky Survey and present the resulting
constraints on both cosmological parameters (fully marginalized over
the uncertainties related to galaxy bias) and the CLF parameters
(fully marginalized over uncertainties in cosmological parameters).

\section*{Acknowledgments}

The work presented in this paper has greatly benefited from
discussions with Matthew Becker, Alexie Leauthaud, Nikhil Padmanabhan,
Eduardo Rozo, Roman Scoccimarro, Jeremy Tinker, Risa Wechsler, Idit
Zehavi and Zheng Zheng. The analysis of numerical simulations used in
this work has been performed on the Joint Fermilab - KICP
Supercomputing Cluster, supported by grants from Fermilab, Kavli
Institute for Cosmological Physics, and the University of
Chicago. FvdB acknowledges support from the Lady Davis Foundation for
a Visiting Professorship at Hebrew University. This research was
supported in part by the National Science Foundation under Grant
No. NSF PHY11-25915 and NSF PHY-0551142.


\label{lastpage}
\end{document}

%% file: macros.tex
\def\beq{\begin{equation}}
\def\eeq{\end{equation}}
\def\barray{\begin{eqnarray}}
\def\earray{\end{eqnarray}}
\def\beqarray{\begin{eqnarray}}
\def\eeqarray{\end{eqnarray}}

\def\rmb{{\rm b}}
\def\rmc{{\rm c}}
\def\rmd{{\rm d}}
\def\rme{{\rm e}}

\def\rmg{{\rm g}}
\def\rmh{{\rm h}}

\def\rmm{{\rm m}}

\def\rmp{{\rm p}}

\def\rms{{\rm s}}
\def\rmt{{\rm t}}

\def\rmx{{\rm x}}
\def\rmy{{\rm y}}

\def\rmD{{\rm D}}

\def\rmP{{\rm P}}

\def\calA{{\cal A}}

\def\calH{{\cal H}}

\def\calN{{\cal N}}

\def\calP{{\cal P}}

\def\calR{{\cal R}}

\def\bk{{\bf k}}

\def\br{{\bf r}}

\def\bx{{\bf x}}
\def\by{{\bf y}}

\newcommand{\etal}{{et al.~}}

\newcommand{\kmsmpc}{\>{\rm km}\,{\rm s}^{-1}\,{\rm Mpc}^{-1}}

\newcommand{\Gpc}{\>{\rm Gpc}}
\newcommand{\Mpc}{\>{\rm Mpc}}

\newcommand{\Msun}{\>{\rm M_{\odot}}}
\newcommand{\Lsun}{\>{\rm L_{\odot}}}

\newcommand{\sigc}{\sigma_{{\rm ln} c}}

\newcommand{\bolds}[1]{{\bf #1}}

\def\gtsima{$\; \buildrel > \over \sim \;$}
\def\ltsima{$\; \buildrel < \over \sim \;$}
\def\prosima{$\; \buildrel \propto \over \sim \;$}
\def\gsim{\lower.7ex\hbox{\gtsima}}
\def\lsim{\lower.7ex\hbox{\ltsima}}
\def\simgt{\lower.7ex\hbox{\gtsima}}
\def\simlt{\lower.7ex\hbox{\ltsima}}
\def\simpr{\lower.7ex\hbox{\prosima}}
\def\la{\lsim}
\def\ga{\gsim}
\def\lta{\la}
\def\gta{\ga}

\newcommand{\apj}{ApJ}
\newcommand{\apjs}{ApJS}

\newcommand{\aap}{A\&A}

\newdimen\hssize
\newdimen\hdsize
